\documentclass[pra,12pt,tightenlines,showpacs,showkeys]{revtex4}
\usepackage{latexsym}
\usepackage{graphicx}
\usepackage{amsmath}
\usepackage{amsbsy}
\usepackage{amsthm}
\usepackage{bbm}
\usepackage{bm}

\newtheorem{lem}{Lemma}

\begin{document}

\title{Entanglement witnesses and geometry of entanglement of two--qutrit states}
\author{Reinhold A. Bertlmann} \email{reinhold.bertlmann@univie.ac.at}
\author{Philipp Krammer} \email{philipp.krammer@univie.ac.at}
\affiliation{Faculty of Physics, University of Vienna, Boltzmanngasse 5, A-1090 Vienna,
Austria}

\begin{abstract}

We construct entanglement witnesses with regard to the geometric
structure of the Hilbert--Schmidt space and investigate the geometry
of entanglement. In particular, for a two--parameter family of
two--qutrit states that are part of the \emph{magic simplex} we
calculate the Hilbert--Schmidt measure of entanglement. We present a
method to detect bound entanglement which is illustrated for a
three--parameter family of states. In this way, we discover new
regions of bound entangled states. Furthermore, we outline how to use
our method to distinguish entangled from separable states.

\end{abstract}


\maketitle

\section{Introduction}

Entanglement is one of the most striking features of quantum theory
and is of capital importance for the whole field of quantum
information theory (see, e.g., Refs.~\cite{bertlmann02a,
bouwmeester00, nielsen00}). The determination whether a given
quantum state is entangled or separable is still an open and
challenging problem, in particular for higher dimensional systems.

For a two--qubit system the geometric structure of the entangled and separable states in
the Hilbert--Schmidt space is very well known. Due to the Peres--Horodecki criterion
\cite{peres96, horodecki96} we know necessary and sufficient conditions for separability.
This case is, however, due to its high symmetry quite unique and even misleading for
conclusions in higher dimensions.

In higher dimensions, the geometric structure of the states is much more complicated and
new phenomena like bound entanglement occur \cite{horodecki01, horodecki97a, horodecki98,
horodecki99c, rains99, baumgartner07}. Although we do not know necessary and sufficient
conditions \`{a} la Peres--Horodecki we can construct an operator
-- entanglement witness -- which provides via an inequality a
criterion for the entanglement of the state \cite{terhal00,
terhal02, horodecki96, bertlmann02}.

In this paper, we use entanglement witnesses to explore the geometric
structure of the two--qutrit states in Hilbert--Schmidt space, i.e.
we geometrically quantify entanglement for special cases, and
present a method to detect bound entanglement. Two qutrits are
states on the $3 \times 3$ dimensional Hilbert space of bipartite
quantum systems. In analogy to the familiar two--qubit case, which
we discuss at the beginning, we introduce a two-- and
three--parameter family of two--qutrit states which are part of the
\emph{magic simplex} of states \cite{baumgartner06, baumgartner07,
baumgartner08} and determine geometrical properties of the states:
For the two--parameter family we quantify the entanglement via the
Hilbert--Schmidt measure, for the three--parameter family we
discover bound entangled states in addition to known ones in the
simplex \cite{horodecki99c, baumgartner06}. Finally, we give a sketch
of how to use our method to construct the shape of the separable
states for the three--parameter family.

\section{Weyl operator basis} \label{secwob}

As \emph{standard matrix basis} we consider the standard matrices,
the $d \times d$ matrices, that have only one entry $1$ and the
other entries $0$ and form an orthonormal basis of the
Hilbert--Schmidt space, which is the space of operators that act on
the states of the Hilbert space ${\cal H}^d$ of dimension $d$. We
write these matrices shortly as operators
\begin{equation} \label{standardmatrices}
    | j \rangle \langle k | \,, \qquad \textrm{with} \quad j,k = 1, \ldots, d
    \,,
\end{equation}
where the matrix representation can be easily obtained in the
standard vector basis $\left\{ | i \rangle \right\}$. Any matrix can
easily be decomposed into a linear combination of matrices
\eqref{standardmatrices}.

The Weyl operator basis (WOB) of the Hilbert--Schmidt space of
dimension $d$ consists of the following $d^2$ operators (see
Ref.~\cite{bertlmann08a})
\begin{equation} \label{defwo}
    U_{nm} \;=\; \sum_{k=0}^{d-1} e^{\frac{2 \pi i}{d}\,kn} \,| k \rangle
    \langle (k+m) \,\textrm{mod}\,d| \qquad n,m = 0,1, \ldots ,d-1
    \,.
\end{equation}
These operators have been frequently used in the literature (see
e.g. Refs.~\cite{narnhofer06, baumgartner06, baumgartner07,
bennett93}), in particular, to create a basis of $d^2$ maximally
entangled qudit states \cite{narnhofer06, werner01, vollbrecht00}.\\

\emph{Example.} In the case of qutrits, i.e. of a three--dimensional
Hilbert space, the Weyl operators \eqref{defwo} have the following
matrix form:
\begin{eqnarray}
    & U_{00} & = \begin{pmatrix}
        1 & 0 & 0 \\
        0 & 1 & 0 \\
        0 & 0 & 1
       \end{pmatrix}, \quad\qquad\qquad\,
    U_{01} = \begin{pmatrix}
        0 & 1 & 0 \\
        0 & 0 & 1 \\
        1 & 0 & 0
        \end{pmatrix}, \quad\qquad\qquad\,
    U_{02} = \begin{pmatrix}
        0 & 0 & 1 \\
        1 & 0 & 0 \\
        0 & 1 & 0
        \end{pmatrix},  \\
    & U_{10} & = \begin{pmatrix}
        1 & 0 & 0 \\
        0 & e^{2 \pi i/3} & 0 \\
        0 & 0 & e^{-2 \pi i/3}
        \end{pmatrix}, \quad
    U_{11} = \begin{pmatrix}
        0 & 1 & 0 \\
        0 & 0 & e^{2 \pi i/3} \\
        e^{-2 \pi i/3} & 0 & 0
        \end{pmatrix}, \quad
    U_{12} = \begin{pmatrix}
        0 & 0 & 1 \\
        e^{2 \pi i/3} & 0 & 0 \\
        0 & e^{-2 \pi i/3} & 0
        \end{pmatrix},  \nonumber\\
    & U_{20} & = \begin{pmatrix}
        1 & 0 & 0 \\
        0 & e^{-2 \pi i/3} & 0 \\
        0 & 0 & e^{2 \pi i/3}
        \end{pmatrix}, \quad
    U_{21} = \begin{pmatrix}
        0 & 1 & 0 \\
        0 & 0 & e^{-2 \pi i/3} \\
        e^{2 \pi i/3} & 0 & 0
        \end{pmatrix}, \quad
    U_{22} = \begin{pmatrix}
        0 & 0 & 1 \\
        e^{-2 \pi i/3} & 0 & 0 \\
        0 & e^{2 \pi i/3} & 0
        \end{pmatrix}. \nonumber
\end{eqnarray}\\

Using the WOB we can decompose quite generally any density matrix in
form of a vector, called \textit{Bloch vector} \cite{bertlmann08a}
\begin{equation} \label{bvwob-d}
    \rho \;=\; \frac{1}{d} \,\mathbbm{1} \,+\, \sum_{n,m=0}^{d-1} b_{nm} U_{nm}
     \;=\; \frac{1}{d} \,\mathbbm{1} \,+\, \vec{b} \cdot \vec{U} \,,
\end{equation}
with $n,m = 0,1, ... ,d-1$ ($b_{00}=0$). The components of the Bloch
vector $\vec{b} = \big(\{b_{nm}\}\big)$ are ordered and given by
$b_{nm} = \textnormal{Tr}\,U_{nm}\,\rho\,$. In general, the
components $b_{nm}$ are complex since the Weyl operators are not
Hermitian and the complex conjugates fulfil the relation
$b_{n\,m}^\ast = e^{-\frac{2 \pi i}{d}\,nm} \, b_{{-n}{-m}}\,$,
which follows easily from definition \eqref{defwo} together with the
hermiticity of $\rho\,$. Note that for $d \geq 3$ not any vector
$\vec{b}$ of complex components is a Bloch vector, i.e. a quantum
state (details can be found in Ref.~\cite{bertlmann08a}).

The standard matrices \eqref{standardmatrices}, on the other hand,
can be expressed by the WOB as \cite{bertlmann08a}
\begin{equation} \label{smwob}
    |j \rangle \langle k | \;=\; \frac{1}{d}\, \sum_{l=0}^{d-1} e^{-\frac{2 \pi i}{d}
    \,lj} \;U_{l\, (k-j)\,\textrm{mod}\,d} \;.
\end{equation}
It allows us to decompose any two--qudit state of a bipartite system
in the $d\times d$ dimensional Hilbert space ${\cal H}^{\,d}_A
\otimes {\cal H}^{\,d}_B$ into the WOB. In particular, we consider
the \emph{isotropic} two--qudit state $\rho_{\alpha}^{(d)}$ which is
defined as follows \cite{horodecki99, rains99, horodecki01} \,:
\begin{equation} \label{rhodefiso}
    \rho_{\alpha}^{(d)} \;=\; \alpha \left| \phi_+^d \right\rangle \left\langle \phi_+^d \right|
    \,+\, \frac{1-\alpha}{d^2}\,\mathbbm{1}\,, \quad \alpha \in \mathbbm{R}\,, \quad
    - \frac{1}{d^2-1} \leq \alpha \leq 1 \;.
\end{equation}
The range of $\alpha$ is determined by the positivity of the state.

The state
\begin{equation} \label{defmaxent}
    \left| \phi^d_+ \right\rangle \;=\; \frac{1}{\sqrt{d}} \,
    \sum_j \left| j \right\rangle \otimes \left| j \right\rangle
\end{equation}
denotes a Bell state, which is maximally entangled, and in terms of
the WOB it is expressed by (see Refs.~\cite{bertlmann08a,
narnhofer06})
\begin{equation} \label{maxentwob2}
    \left| \phi_+^d \right\rangle \left\langle \phi_+^d \right| \;=\;
    \frac{1}{d^2} \,\mathbbm{1} \otimes \mathbbm{1} \,+\, \frac{1}{d^2} \, U \,,
\end{equation}
with
\begin{equation} \label{defu}
    U \;:=\; \sum_{l,m = 0}^{d-1} U_{lm} \otimes
        U_{-lm} \,, \qquad (l,m) \neq (0,0) \,.
\end{equation}
The negative values of the index $l$ have to be considered as $mod \ d\,$.

Then the isotropic two--qudit state follows from Eq. \eqref{maxentwob2}:
\begin{equation} \label{isowob}
    \rho^{(d)}_\alpha \;=\; \frac{1}{d^2} \,\mathbbm{1} \otimes \mathbbm{1}
    \,+\, \frac{\alpha}{d^2} \, U \,.
\end{equation}
We call decomposition \eqref{isowob} \textit{Bloch vector form} of
the density matrix.

\section{Entanglement witness --- Hilbert--Schmidt measure}\label{secapp}

\subsection{Definitions} \label{def-entwitness}

The \emph{Hilbert--Schmidt} (HS) \emph{measure} of entanglement
\cite{witte99,ozawa00,bertlmann02} is defined as the minimal HS
distance of an entangled state $\rho_{\rm{ent}}$ to the set of
separable states $S$,
\begin{equation} \label{defhs}
    D(\rho_{\rm{ent}}) \;:=\; \min_{\rho \in S} \left\| \rho - \rho_{\rm{ent}} \right\|
    \;=\; \left\| \rho_0 - \rho_{\rm{ent}} \right\| \,,
\end{equation}
where $\rho_0$ denotes the nearest separable state, the minimum of the HS distance.

An \emph{entanglement witness} $A$ is a Hermitian operator that
``detects'' the entanglement of a state $\rho_{\rm ent}$ via
inequalities 
\cite{horodecki96, terhal00, terhal02, bertlmann02, bertlmann08a}
\begin{eqnarray} \label{defentwit}
    \left\langle \rho_{\rm ent},A \right\rangle \;=\; \textnormal{Tr}\, \rho_{\rm ent} A
    & \;<\; & 0 \,,\nonumber\\
    \left\langle \rho,A \right\rangle = \textnormal{Tr}\, \rho A & \;\geq\; & 0 \qquad
    \forall \rho \in S \,.
\end{eqnarray}
An entanglement witness is ``optimal'', denoted by $A_{\rm{opt}}\,$, if apart from
Eq.~(\ref{defentwit}) there exists a separable state $\rho_0 \in S$ such that
\begin{equation}
    \left\langle \rho_0 ,A_{\rm{opt}} \right\rangle \;=\; 0 \,.
\end{equation}
The operator $A_{\rm{opt}}$ defines a tangent plane to the set of
separable states $S$ and can be constructed in the following way
\cite{bertlmann02, bertlmann05} :
\begin{equation} \label{entwitmaxviolation}
    A_{\rm{opt}} \;=\; \frac{\rho_0 - \rho_{\rm ent} \,-\, \left\langle \rho_0 ,
    \rho_0 - \rho_{\rm ent} \right\rangle \mathbbm{1}}{\left\| \rho_0 - \rho_{\rm ent}
    \right\|} \;.
    \end{equation}
The term $\left\| \rho_0 - \rho_{\rm ent} \right\|$ is a
normalization factor and is included for a convenient scaling of the
inequalities \eqref{defentwit} only. Thus, the calculation of the
optimal entanglement witness $A_{\rm{opt}}$ to a given entangled
state $\rho_{\rm ent}$ reduces to the determination of the nearest
separable state $\rho_0\,$. In special cases, we are able to find
$\rho_0$ by a \emph{guess method} \cite{bertlmann05}, what we
demonstrate in this article by working with our Lemmas, but in
general its detection is quite a difficult task.

\subsection{Two--parameter entangled states --- qubits}\label{secqubits}

\begin{figure}
  \begin{centering}
    \includegraphics[width=0.5\textwidth]{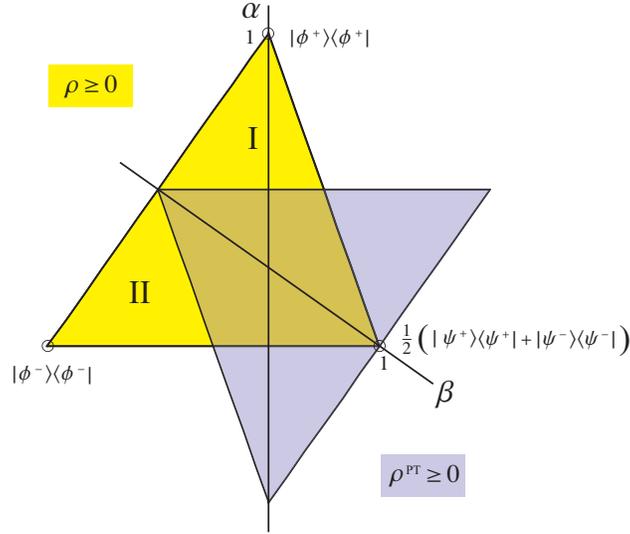}\\
    \caption{Illustration of the two--qubit states $\rho_{\alpha, \beta}$ \eqref{rhoqubit}
    and their partial transpositions $\rho^{\rm{PT}}$.}
    \label{figqubit}
  \end{centering}
\end{figure}

It is quite illustrative to start with the familiar case of a
two--qubit system to gain some intuition of the geometry of the
states in the HS space. We aim to determine the entanglement witness
and the HS measure of entanglement for the following two--qubit
states which are a particular mixture of the Bell states $| \psi^-
\rangle, | \psi^+ \rangle, | \phi^- \rangle, | \phi^+ \rangle$:
\begin{equation} \label{rhoqubit}
    \rho_{\alpha, \beta} \;=\; \frac{1-\alpha-\beta}{4} \,\mathbbm{1} \,+\,
    \alpha | \phi^+ \rangle \langle \phi^+ | \,+\, \frac{\beta}{2}
    \left( | \psi^+ \rangle \langle \psi^+ | \,+\, | \psi^- \rangle
    \langle \psi^- | \right) \,.
\end{equation}
The states \eqref{rhoqubit} are characterized by the two parameters $\alpha$ and $\beta$
and we will refer to the states as the \emph{two--parameter states}. Of course, the
positivity requirement $ \rho_{\alpha, \beta} \geq 0$ constrains the possible values of
$\alpha$ and $\beta$, namely
\begin{equation}
    \alpha \;\leq\; - \beta +1, \quad\; \alpha \;\geq\; \frac{1}{3} \beta - \frac{1}{3},
    \quad\; \alpha \;\leq\; \beta+1 \,,
\end{equation}
which geometrically corresponds to a triangle, see Fig.~\ref{figqubit}.

According to Peres \cite{peres96} and the Horodeckis
\cite{horodecki96} the separability of the states is determined by
the \emph{positive partial transposition criterion}, which
says that a separable state has to stay positive under partial
transposition (PPT). For dimensions $2 \times 2$ and $2 \times 3$ the
criterion is necessary and sufficient \cite{horodecki96}, thus any
PPT state is separable for these dimensions. States \eqref{rhoqubit}
which are positive under partial transposition have the following
constraints:
\begin{equation}
    \alpha \;\geq\; \beta -1, \quad\; \alpha \;\leq\; \frac{1}{3} \beta + \frac{1}{3},
    \quad\; \alpha \geq -\beta-1 \,,
\end{equation}
and correspond to the rotated triangle; then the overlap, a rhombus, represents the
separable states, see Fig.~\ref{figqubit}.

In the illustration of Fig.~\ref{figqubit} the orthogonal lines are
indeed orthogonal in HS space. Therefore, the coordinate axes for the
parameter $\alpha$ and $\beta$ are necessarily non--orthogonal. In
particular, the $\alpha$ axis has to be orthogonal to the boundary
line $\alpha = 1/3(\beta - 1)$, and the $\beta$ axis has to be
orthogonal to $\alpha = \beta+1$.

\begin{figure}
  \begin{centering}
    \includegraphics[width=0.5\textwidth]{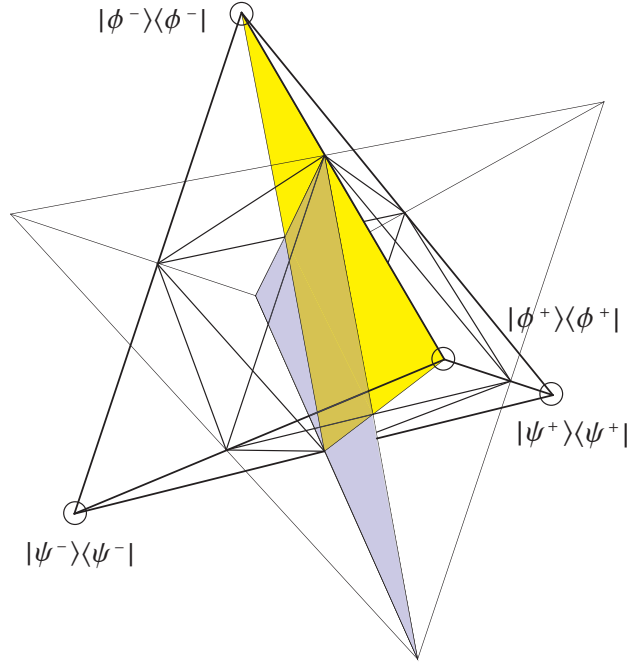}\\
    \caption{Location of the plane of states $\rho_{\alpha, \beta}$
    \eqref{rhoqubit} in the tetrahedron formed by the Bell states.}
    \label{figtetrahedron}
  \end{centering}
\end{figure}

The two--parameter states $\rho_{\alpha, \beta}$ define a plane in
the HS space. It is quite illustrative to see how this plane is
located in the three--dimensional simplex formed by states that are
mixtures of the the four two-qubit Bell states. The simplex
represents a tetrahedron due to the positivity conditions of the
density matrix \cite{bertlmann02, vollbrecht01, horodecki96b}.
Applying PPT, the tetrahedron is rotated producing an intersection
-- a double pyramid -- which corresponds to the separable states. An
illustration of the described features is given in
Fig.~\ref{figtetrahedron}.

To calculate the HS measure \eqref{defhs} for the two--parameter qubit state
\eqref{rhoqubit} we express the state in terms of the Pauli matrix basis
\begin{equation} \label{rhoqubitpauli}
    \rho_{\alpha,\beta} \;=\; \frac{1}{4} \left( \mathbbm{1} \,+\, \alpha
    \left( \sigma_1 \otimes \sigma_1 \,-\, \sigma_2 \otimes \sigma_2
    \right) \,+\, \left( \alpha - \beta \right)
    \sigma_3 \otimes \sigma_3 \right) \,,
\end{equation}
where we have used the well--known Pauli matrix decomposition of the Bell states (see, e.g.,
Ref.~\cite{bertlmann02}).\\

In order to determine the HS measure for the entangled two--parameter states
$\rho_{\alpha,\beta}^{\rm{ent}}$ we have to find the nearest separable states, which is
usually the most difficult task to perform in this context. In Ref.~\cite{bertlmann05} a
Lemma is presented to check if a particular separable state is indeed the nearest
separable state to a given entangled one.
\begin{lem} \label{lemsepablenearest}
    A state $\tilde{\rho}$ is equal to the nearest separable state $\rho_0$
    if and only if the operator
    \begin{equation} \label{ctilde}
    \tilde{C} \;=\; \frac{\tilde{\rho} - \rho_{\rm ent} \,-\, \left\langle \tilde{\rho} ,
    \tilde{\rho} - \rho_{\rm ent} \right\rangle \mathbbm{1}}{\left\| \tilde{\rho} -
    \rho_{\rm ent} \right\|}
\end{equation}
is an entanglement witness.
\end{lem}
Lemma~\ref{lemsepablenearest} probes if a guess $\tilde{\rho}$ is indeed correct for the
nearest separable state. If this is the case, operator $\tilde{C}$ represents the optimal entanglement witness $A_{\rm{opt}}$ \eqref{entwitmaxviolation}.\\

Lemma~\ref{lemsepablenearest} is used here in the following way.
First, for any entangled two-parameter state \eqref{rhoqubit} we
calculate the separable state that has the nearest Euclidean
distance in the geometric picture (Fig.~\ref{figqubit}) and call
this state $\tilde{\rho}$. But since the regarded picture does not
represent the full state space (e.g., states containing terms like
$a_i\,\sigma^i \otimes \mathbbm{1}$ or $b_i\,\mathbbm{1} \otimes
\sigma^i$ are not contained on the picture), we have to use
Lemma~\ref{lemsepablenearest} to check if the estimated state
$\tilde{\rho}$ is indeed the nearest separable state $\rho_0$.

\subsubsection{Region I}

Let us consider first the entangled states located in the triangle region that includes
the Bell state $| \phi^+ \rangle\,$, i.e. Region I in Fig.~\ref{figqubit}. An entangled
state in Region I is characterized by points, i.e. by the parameter pair
($\alpha$,$\beta$), constrained by
\begin{equation}
    \alpha \;\leq\; \beta + 1, \quad\; \alpha \;\leq\; - \beta +1, \quad\; \alpha \;>\;
    \frac{1}{3}\beta + \frac{1}{3} \,.
\end{equation}
The point in the separable region of Fig.~\ref{figqubit} that is nearest (in the
Euclidean sense) to the point ($\alpha$,$\beta$) is given by ($\frac{1}{3} +
\frac{1}{3}\beta$,$\beta$), which corresponds to the state
\begin{equation} \label{rhosep-qubitpauli}
    \tilde{\rho}_\beta \;=\; \frac{1}{4} \left( \mathbbm{1} \,+\,
    \frac{1+\beta}{3} \,
    \left( \sigma_1 \otimes \sigma_1 \,-\, \sigma_2 \otimes \sigma_2
    \right) \,+\, \frac{1-2\beta}{3} \, \sigma_3 \otimes \sigma_3 \right) \,.
\end{equation}
For the difference of the ``nearest Euclidean separable'' and the
entangled state, we obtain
\begin{equation} \label{rhominusqubit}
    \tilde{\rho}_\beta \,-\, \rho^{\rm{ent}}_{\alpha, \beta} \;=\; \frac{1}{4}
    \left( \frac{1+\beta}{3} \,-\, \alpha \right) \Sigma \,,
\end{equation}
where $\Sigma$ is defined by
\begin{equation}
    \Sigma \;:=\; \sigma_1 \otimes \sigma_1 \,-\, \sigma_2 \otimes \sigma_2
    \,+\, \sigma_3 \otimes \sigma_3 \,.
\end{equation}
Using the norm $\| \Sigma \| = 2 \sqrt{3}$ we gain the HS distance
\begin{equation} \label{hsqubit}
    \| \tilde{\rho}_\beta - \rho_{\alpha,\beta}^{\rm{ent}} \| \;=\;
    \frac{\sqrt{3}}{2} \left( \alpha - \frac{1}{3} - \frac{1}{3}
    \beta \right) \,.
\end{equation}
To check whether the state $\tilde{\rho}_\beta$ coincides with the nearest separable
state $\rho_{0; \beta}$ in the sense of the HS measure of entanglement \eqref{defhs}
(which has to take into account the \emph{whole} set of separable states), we have to
test -- according to Lemma~\ref{lemsepablenearest} -- whether the operator
\begin{equation}
    \tilde{C} \;=\; \frac{ \tilde{\rho}_\beta - \rho^{\rm{ent}}_{\alpha,\beta} \,-\,
    \langle \tilde{\rho}_\beta, \tilde{\rho}_\beta -
    \rho^{\rm{ent}}_{\alpha,\beta} \rangle \,\mathbbm{1} }{ \| \tilde{\rho}_\beta -
    \rho_{\alpha,\beta}^{\rm{ent}} \| }
\end{equation}
is an entanglement witness. Remember that any entanglement witness $A$ that detects the
entanglement of a state $\rho^{\rm{ent}}$ has to satisfy the inequalities
\eqref{defentwit}.

We calculate
\begin{equation}
    \langle \tilde{\rho}_\beta , \tilde{\rho}_\beta - \rho^{\rm{ent}}_{\alpha,\beta}
    \rangle \;=\; \textnormal{Tr} \,\tilde{\rho}_\beta ( \tilde{\rho}_\beta -
    \rho^{\rm{ent}}_{\alpha,\beta} ) \;=\; -\frac{1}{4} \left( \alpha - \frac{1}{3} -
    \frac{1}{3} \beta \right)
\end{equation}
and use Eqs.~\eqref{rhominusqubit} and \eqref{hsqubit} to determine the operator
$\tilde{C}$ for the considered case
\begin{equation} \label{concretectilde}
    \tilde{C} \;=\; \frac{1}{2 \sqrt{3}} \,\left( \mathbbm{1} \,-\, \Sigma \right) \,.
\end{equation}
Then we find
\begin{equation} \label{Centinequal}
    \langle \rho^{\rm{ent}}_{\alpha,\beta}, \tilde{C} \rangle \;=\;
    - \frac{\sqrt{3}}{2} \left( \alpha - \frac{1}{3} - \frac{1}{3}
    \beta \right) \;<\; 0 \,,
\end{equation}
since the entangled states in the considered Region I satisfy the constraint $\alpha >
\frac{1}{3}{\beta} + \frac{1}{3}\,$. Thus, the first condition of inequalities
\eqref{defentwit} is fulfilled.

Actually, condition \eqref{Centinequal} is just a consistency check for the correct
calculation of operator $\tilde{C}$ since by construction of $\tilde{C}$ we always have $
\langle \rho^{\rm{ent}}, \tilde{C} \rangle \;=\; - \| \tilde{\rho} - \rho^{\rm{ent}} \|
\;<\; 0 \,$. Thus more important is the test of the second condition of inequalities
\eqref{defentwit} and in order to do it we need the following Lemma.
\begin{lem} \label{lemqubit}
    For any Hermitian operator $C$ on a Hilbert space of dimension $2 \times 2$ that is of the form
    \begin{equation} \label{lemqubitc}
        C \;=\; a\, \left(\mathbbm{1} \,+\, \sum_{i=1}^3 c_i \,\sigma_i \otimes
        \sigma_i \right)
        \qquad a \in \mathbbm{R^+}, \ c_i \in \mathbbm{R}
    \end{equation}
    the expectation value for all separable states is positive,
    \begin{equation}
        \langle \rho , C \rangle \,\geq\, 0 \quad \forall \,\rho \in S \,,
        \quad \mbox{\textrm{if}} \quad |c_{i}| \,\leq\, 1 \quad \forall \, i \,.
    \end{equation}
\end{lem}

\noindent\emph{Proof.} Any separable state $\rho$ is a convex combination of product
states and thus a separable two--qubit state can be written as (see
Refs.~\cite{bertlmann02, bertlmann05})
\begin{eqnarray} \label{seppauli}
    & \rho \;=\; \sum_k p_k \, \frac{1}{4} \left(
        \mathbbm{1} \otimes \mathbbm{1}
        \,+\, \sum_i n_i^k\,\sigma_i \otimes \mathbbm{1} \,+\, \sum_j m_j^k\,
        \mathbbm{1} \otimes \sigma_j
        \,+\, \sum_{i,j} n_i^k m_j^k \,\sigma_i \otimes \sigma_j \right)\,, & \nonumber\\
    & \textrm{with}\quad n_i^k, m_i^k \in \mathbbm{R}\,, \; \left| \vec{n}^k \right| \leq
        1\,, \left| \vec{m}^k \right| \leq 1 \,, \ \ p_k \geq 0, \, \sum_k p_k = 1
        \,, &
\end{eqnarray}
where $|\vec{n}^k|^2 = \sum_i \left( n^k_i \right)^2$. Performing the trace, we obtain
\begin{equation}
    \langle \rho , C \rangle \;=\; \textnormal{Tr} \, \rho \, C \;=\; \sum_k p_k
    \,a \left( 1 \,+\, \sum_i c_i \, n_i^k m_i^k \right) \,,
\end{equation}
and using the restriction $|c_{i}| \,\leq\, 1 \;\; \forall \, i$ we have
\begin{equation}
    \left| \sum_i c_i \, n_i^k m_i^k \right| \;\leq\; \sum_i |n_i^k| |m_i^k| \;\leq\; 1 \,,
\end{equation}
and since the convex sum of positive terms stays positive we get $\langle \rho , C
\rangle \;\geq\; 0 \;\; \forall \rho \in S \,. \ \Box$\\

Since the operator $\tilde{C}$ \eqref{concretectilde} is of the form \eqref{lemqubitc} we
can use Lemma~\ref{lemqubit} to verify
\begin{equation}
    \langle \rho , \tilde{C} \rangle \;\geq\; 0 \qquad \forall \rho \in S \,.
\end{equation}
Therefore, $\tilde{C}$ \eqref{concretectilde} is indeed an entanglement witness and
$\tilde{\rho}_\beta$ is the nearest separable state $\tilde{\rho}_\beta = \rho_{0; \,
\beta}$ for the entangled states $\rho_{\alpha,\beta}^{\rm{ent}}$ in Region I.\\

Finally, we find for the HS measure of the states in Region I
\begin{equation} \label{hsqubitreal}
    D^I(\rho_{\alpha, \beta}^{\rm{ent}}) \;=\; \| \rho_{0; \, \beta} \,-\,
    \rho_{\alpha,\beta}^{\rm{ent}} \| \;=\;
    \frac{\sqrt{3}}{2} \left( \alpha - \frac{1}{3} - \frac{1}{3} \beta \right) \,.
\end{equation}
Note that the HS measure \eqref{hsqubitreal} is equal to the maximal
violation of the entanglement witness inequality \eqref{defentwit}
as it is shown in detail in
Refs.~\cite{bertlmann02,bertlmann05,bertlmann08a}.

\subsubsection{Region II}

It remains to determine the HS measure for the entangled states
$\rho_{\alpha,\beta}^{\rm{ent}}$ located in the triangle region that includes the Bell
state $| \phi^- \rangle\,$, i.e. Region II in Fig.~\ref{figqubit}. Here, the entangled
states are characterized by points $(\alpha, \beta)\,$, where the parameters are
constrained by
\begin{equation}
    \alpha \;\leq\; \beta + 1, \quad\; \alpha \;\geq\; \frac{1}{3} \beta -
    \frac{1}{3}, \quad\; \alpha \;<\; - \beta - 1 \,.
\end{equation}
The states in the separable region of Fig.~\ref{figqubit} that are nearest to the
entangled states $(\alpha, \beta)$ in Region II are called $\tilde{\rho}_{\alpha, \beta}$
and characterized by the points
\begin{equation}
\left(
    \begin{array}{c}
    \tilde{\alpha} \\
    \tilde{\beta}
    \end{array}
\right) \;=\; \left(
    \begin{array}{c}
    1/3 \,\left ( -1 + 2 \alpha - \beta \right) \\
    1/3 \,\left ( -2 - 2 \alpha + \beta \right)
    \end{array}
\right) \,.
\end{equation}
The necessary quantities for calculating the operator $\tilde{C}$
are
\begin{equation}
    \tilde{\rho}_{\alpha, \beta} - \rho_{\alpha, \beta}^{\rm{ent}} \;=\;
    - \frac{1}{12} \left( \alpha + 1 + \beta \right) \left( \sigma_1
    \otimes \sigma_1 - \sigma_2 \otimes \sigma_2 - \sigma_3 \otimes
    \sigma_3 \right) \,,
\end{equation}
\begin{equation}
    \| \tilde{\rho}_{\alpha, \beta} - \rho_{\alpha, \beta}^{\rm{ent}} \| \;=\;
    \frac{1}{2\sqrt{3}} \left( - \alpha - 1 - \beta \right) \,,
\end{equation}
\begin{equation}
    \langle \tilde{\rho}_{\alpha, \beta} , \tilde{\rho}_{\alpha, \beta}^{\rm{ent}} -
    \rho_{\alpha,\beta} \rangle \;=\; \frac{1}{12} \left( \alpha + 1 + \beta \right) \,,
\end{equation}
so that $\tilde{C}$ is expressed by
\begin{equation} \label{concretectilde2}
    \tilde{C} \;=\; \frac{1}{2\sqrt{3}} \left( \mathbbm{1} \,+\, \sigma_1
    \otimes \sigma_1 \,-\, \sigma_2 \otimes \sigma_2 \,-\, \sigma_3 \otimes
    \sigma_3 \right) \,.
\end{equation}
To test $\tilde{C}$ for being an entanglement witness, we need to check the first
condition of inequalities \eqref{defentwit}; we get
\begin{equation}
    \langle \rho_{\alpha, \beta}^{\rm{ent}} , \tilde{C} \rangle \;=\;
    \frac{1}{2\sqrt{3}} \left( \alpha + 1 + \beta \right) \;<\; 0
\end{equation}
as expected. Since operator $\tilde{C}$ \eqref{concretectilde2} is of the form
\eqref{lemqubitc} we apply Lemma~\ref{lemqubit} and obtain for the separable states
\begin{equation}
    \langle \rho , \tilde{C} \rangle \;\geq\; 0 \qquad \forall \rho \in S \,.
\end{equation}
Therefore, also in Region II, operator $\tilde{C}$ \eqref{concretectilde2} is indeed an
entanglement witness and $\tilde{\rho}_{\alpha,\beta}$ is the nearest separable state
$\tilde{\rho}_{\alpha,\beta} = \rho_{0; \, \alpha \beta}$ for the entangled states
$\rho_{\alpha,\beta}^{\rm{ent}}$.\\

For the HS measure of the states in Region II, we find
\begin{equation}
    D^{II}(\rho_{\alpha,\beta}^{\rm{ent}}) \;=\; \| \rho_{0; \, \alpha, \beta} \,-\,
    \rho_{\alpha,\beta}^{\rm{ent}} \| \;=\; \frac{1}{2\sqrt{3}}
    \left( - \alpha - 1 - \beta \right) \,.
\end{equation}

\subsection{Two--parameter entangled states --- qutrits}\label{secqutrits}

The procedure of determining the geometry of separable and entangled
states discussed in Sec.~\ref{secqubits} can be generalized to
higher dimensions, e.g. for two--qutrit states. Let us first notice
how to generalize the concept of a maximally entangled Bell basis to
higher dimensions. A basis of maximally entangled two--qudit states
can be attained by starting with the maximally entangled qudit state
$| \phi_+^d \rangle$ \eqref{defmaxent} and constructing the other
$d^2-1$ states in the following way:
\begin{equation}
    | \phi_i \rangle \;=\; \tilde{U}_i \otimes \mathbbm{1} \,| \phi_+^d \rangle
    \qquad i=1,2, \ldots ,d^2-1 \,,
\end{equation}
where $\{ \tilde{U}_i \}$ represents an orthogonal matrix basis of \emph{unitary}
matrices and $\tilde{U}_0$ usually denotes the identity matrix $\mathbbm{1}$ (see Refs.
\cite{vollbrecht00, werner01}).

A reasonable choice of the basis of unitary matrices is the WOB (see
Sec.~\ref{secwob}). Such a construction has been proposed in
Ref.~\cite{narnhofer06}. Then we set up the following $d^2$
projectors onto the maximally entangled states -- the Bell states:
\begin{equation} \label{projbell}
    P_{nk} \;:=\; (U_{nk} \otimes \mathbbm{1}) \,| \phi_+^d \rangle \langle \phi_+^d
    |\, (U_{nk}^\dag \otimes \mathbbm{1}) \qquad n,k = 0,1, \ldots ,d-1
    \,.
\end{equation}
In case of qutrits ($d=3$), mixtures of the nine Bell projectors
\eqref{projbell} form an eight--dimensional simplex which is the higher
dimensional analogue of the three--dimensional simplex, the tetrahedron
for qubits, see Fig.~\ref{figtetrahedron}. This eight--dimensional
simplex has a very interesting geometry concerning separability and
entanglement (see Refs.~\cite{baumgartner06, baumgartner07,
baumgartner08}). Due to its high symmetry inside
-- named therefore the \emph{magic simplex} by the authors of
Ref.~\cite{baumgartner06} -- it is enough to consider certain mixtures of Bell states
which form equivalent classes concerning their geometry.

We can express the Bell projectors as Bloch vectors by using the Bloch vector form
\eqref{maxentwob2} of $P_{00} := | \phi_+^d \rangle \langle \phi_+^d |$ and the relations
(indices have to be taken $mod \, d$) \cite{narnhofer06}
\begin{eqnarray}
    U_{nm}^\dag & \;=\; & e^{\frac{2 \pi i}{d}nm} \,U_{-n \, -m} \,, \\
    U_{nm}U_{lk} & \;=\; & e^{\frac{2 \pi i}{d}ml} \,U_{n+l \, m+k} \,.
\end{eqnarray}
It provides for the Bell projectors the Bloch form
\begin{equation} \label{projbloch}
    P_{nk} \;=\; \frac{1}{d^2} \sum_{m,l=0}^{d-1} e^{\frac{2 \pi
    i}{d}(kl-nm)} \,U_{lm} \otimes U_{-lm} \,.
\end{equation}
We are interested in the following two--parameter states of two qutrits as a
generalization of the qubit case, Eq.~\eqref{rhoqubit},
\begin{equation} \label{rhoqutrit}
    \rho_{\alpha, \beta} \;=\; \frac{1- \alpha -\beta}{9} \,\mathbbm{1} \,+\,
    \alpha \,P_{00} \,+\, \beta \frac{1}{2}
    \left( P_{10} + P_{20} \right)\,.
\end{equation}
According to Ref.~\cite{baumgartner06} the Bell states represent
points in a discrete phase space. The indices $n,k$ can be
interpreted as ``quantized'' position coordinate and momentum,
respectively. The Bell states $P_{00},P_{10}$ and $P_{20}$ lie on a
line in this phase space picture of the maximally entangled states,
they exhibit the same geometry as other lines since each line can be
transformed into another one.

\begin{figure}
  \begin{centering}
    \includegraphics[width=0.5\textwidth]{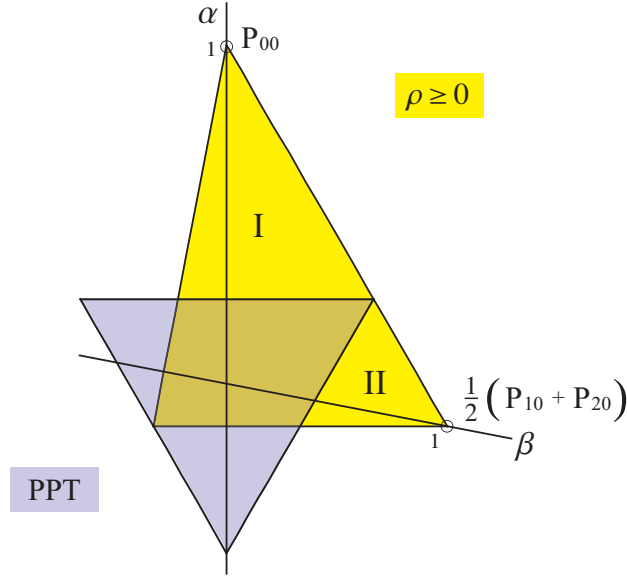}\\
    \caption{Illustration of the two--qutrit states $\rho_{\alpha, \beta}$
    \eqref{rhoqutrit} and their partial transposition. The regions I and II label
    the regions where the states are entangled, they are PPT and separable in the overlap
    with the region of PPT points. The PPT points become semipositive under partial
    transposition.}
    \label{figqutrit}
  \end{centering}
\end{figure}

Inserting the Bloch vector form of $P_{00},P_{10}$ and $P_{20}$ \eqref{projbloch} we find
the Bloch vector expansion of the two--parameter states \eqref{rhoqutrit}
\begin{equation} \label{rhoqutritweyl}
    \rho_{\alpha, \beta} \;=\; \frac{1}{9} \left( \mathbbm{1} \,+\, \left(\alpha -
    \frac{\beta}{2} \right) U_1 \,+\, \left( \alpha + \beta \right)U_2 \right) \,,
\end{equation}
where we defined
\begin{eqnarray} \label{defu1u2}
    U_1 & \;:=\; & U_{01} \otimes U_{01} + U_{02} \otimes U_{02} + U_{11}
    \otimes U_{-11} + U_{12} \otimes U_{-12} + U_{21} \otimes U_{-21}
    + U_{22} \otimes U_{-22} \,, \nonumber\\
    U_2 & \;:=\; & U^I_2 + U^{II}_2  \qquad\mbox{with}\quad
    U^I_2 := U_{10} \otimes U_{-10} \,,\quad U^{II}_2 := U_{20} \otimes U_{-20} \,.
\end{eqnarray}
The constraints for the positivity requirement ($\rho_{\alpha, \beta} \geq 0$) are
\begin{equation}
    \alpha \;\leq\; \frac{7}{2} \beta +1, \quad\; \alpha \;\leq\; -\beta +1, \quad\;
    \alpha \;\geq\; \frac{\beta}{8} - \frac{1}{8} \,,
\end{equation}
and for the PPT
\begin{equation}
    \alpha \;\leq\; - \beta -\frac{1}{2}, \quad\; \alpha \;\geq\; \frac{5}{4}\beta -
    \frac{1}{2}, \quad\; \alpha \;\leq\; \frac{\beta}{8} + \frac{1}{4} \,.
\end{equation}
The Euclidean picture representing the HS space geometry of states \eqref{rhoqutrit} is
shown in Fig.~\ref{figqutrit}. The parameter coordinate axes are chosen non--orthogonal
such that they become orthogonal to the boundary lines of the positivity region, $\alpha
= \frac{\beta}{8} - \frac{1}{8}$ and $\alpha = \frac{7}{2} \beta +1\,$, in order to
reproduce the symmetry of the magic simplex.
It is shown in Ref.~\cite{baumgartner06} that the PPT states
$\rho_{\alpha, \beta}$ are all separable states, thus there exist no
PPT entangled or \textit{bound entangled} states of the form
\eqref{rhoqutrit}. \textit{Bound entanglement} we detect for states
that need more than two parameters in the Bell state expansion (see
Secs.~\ref{secapplmethod} and \ref{secmorebe}).

To find the HS measure for the entangled two--parameter two--qutrit states we apply the
same procedure as in Sec.~\ref{secqubits}: we determine the states that are the nearest
separable ones in the Euclidean sense of Fig.~\ref{figqutrit} and use
Lemma~\ref{lemsepablenearest} to check whether these are indeed the nearest separable
ones with respect to the whole state space (for other approaches see, e.g.,
Refs.~\cite{verstraete02, cao07}).

\subsubsection{Region I}

First, we consider Region I in Fig.~\ref{figqutrit}, i.e., the triangle region of
entangled states around the $\alpha$-axis, constrained by the parameter values
\begin{equation}
    \alpha \;\leq\; \frac{7}{2} \beta +1 , \quad\; \alpha \leq -\beta +1 , \quad\;
    \alpha > \frac{\beta}{8} + \frac{1}{4} \,.
\end{equation}
In the Euclidean picture, the point that is nearest to point $(\alpha, \beta)$ in this
region is given by $(\frac{1}{4}+\frac{1}{8}\beta, \beta)$, which corresponds to the
separable two--qutrit state
\begin{equation}
    \tilde{\rho}_\beta \;=\; \frac{1}{9} \left( \mathbbm{1} \,+\, \left(
    \frac{1}{4} - \frac{3}{8}\beta \right) U_1 \,+\, \left( \frac{1}{4}
    + \frac{9}{8}\beta \right) U_2 \right) \,,
\end{equation}
with $U_1$ and $U_2$ defined in Eq.~\eqref{defu1u2}.

For the difference of ``nearest Euclidean separable'' and entangled state, we find
\begin{equation} \label{rhominusqutrit}
    \tilde{\rho}_\beta - \rho^{\rm{ent}}_{\alpha, \beta} \;=\;
    \frac{1}{9} \left( \frac{1}{4} + \frac{1}{8}\beta - \alpha \right) U \,,
\end{equation}
where $U = U_1 + U_2$ (defined in Eq.~\eqref{defu}). Using for the
norm $\| U \| = 3 \sqrt{8} = 6 \sqrt{2}$ we gain the HS distance
\begin{equation} \label{hsqutrit}
    \| \tilde{\rho}_\beta - \rho_{\alpha,\beta}^{\rm{ent}} \| \;=\;
    \frac{2\sqrt{2}}{3} \left( \alpha - \frac{1}{4} - \frac{1}{8}
    \beta \right) \,.
\end{equation}
It remains to calculate
\begin{equation}
    \langle \tilde{\rho}_\beta , \tilde{\rho}_\beta - \rho_{\alpha,\beta}
    \rangle \;=\; \textnormal{Tr} \,\tilde{\rho}_\beta ( \tilde{\rho}_\beta -
    \rho_{\alpha,\beta} ) \;=\; -\frac{2}{9} \left( \alpha - \frac{1}{4} - \frac{1}{8}
    \beta \right)
\end{equation}
to set up the operator
\begin{equation} \label{ctildequtrit}
    \tilde{C} \;=\; \frac{ \tilde{\rho}_\beta - \rho^{\rm{ent}}_{\alpha,\beta} \,-\,
    \langle \tilde{\rho}_\beta, \tilde{\rho}_\beta - \rho^{\rm{ent}}_{\alpha,\beta}
    \rangle \,\mathbbm{1} }{ \| \tilde{\rho}_\beta - \rho_{\alpha,\beta}^{\rm{ent}} \| } \;=\;
    \frac{1}{6\sqrt{2}} \,\left( 2\,\mathbbm{1} \,-\, U \right) \,.
\end{equation}
We test now whether it represents an entanglement witness, i.e., whether $\tilde{C}$
\eqref{ctildequtrit} satisfies the inequalities \eqref{defentwit}. As expected, we find
\begin{equation}
    \langle \rho^{\rm{ent}}_{\alpha,\beta}, \tilde{C} \rangle \;=\; - \frac{2\sqrt{2}}{3}
    \left( \alpha - \frac{1}{4} - \frac{1}{8} \beta \right) \;<\; 0 \,.
\end{equation}
To check the second condition of inequalities \eqref{defentwit} we set up the following
Lemma, similar to Lemma~\ref{lemqubit}.
\begin{lem} \label{lemqutrit}
    For any Hermitian operator $C$ of a bipartite Hilbert-Schmidt
    space of dimension $d \times d$ that is of the form
    \begin{equation} \label{lemqutritc}
        C \;=\; a \left( (d-1) \, \mathbbm{1}_{\rm{d^2}} \,+\, \sum_{n,m=0}^{d - 1} c_{nm}
            \,U_{nm} \otimes U_{-nm} \right), \quad a \in \mathbbm{R}^+, \
            c_{nm} \in \mathbbm{C}
    \end{equation}
    the expectation value for all separable states is positive,
    \begin{equation} \label{ewineqc}
        \langle \rho , C \rangle \,\geq\, 0 \quad \forall \, \rho \in S \,,
        \quad \mbox{\textrm{if}} \quad |c_{nm}| \,\leq\, 1 \quad \forall \, n,m \,.
    \end{equation}
\end{lem}

\noindent\emph{Proof.} Any bipartite separable state can be decomposed into Weyl
operators as
\begin{align} \label{sepweyl}
    \rho \;=\; & \sum_k p_k \ \frac{1}{d^2} \Big( \mathbbm{1_{\rm{d}}} \otimes
        \mathbbm{1_{\rm{d}}}
        \,+\, \sum_{n,m=0}^{d - 1} \sqrt{d-1}\, n_{nm}^k \,U_{nm} \otimes
        \mathbbm{1_{\rm{d}}}
        \,+\, \sum_{l,k=0}^{d - 1} \sqrt{d-1}\, m_{lk}^k \,\mathbbm{1_{\rm{d}}}
        \otimes U_{lk} \nonumber\\
    & + \sum_{n,m,l,k=0}^{d_1 - 1} (d-1)\, n_{nm}^k m_{lk}^k U_{nm} \otimes U_{lk} \Big) \,,
        \nonumber\\
    & n_{nm}^k, m_{lk}^k \in \mathbbm{C}\,, \;\quad \left| \vec{n}^k \right| \leq
        1\,,\quad \left| \vec{m}^k \right| \leq 1 \,, \quad \ p_k \geq 0, \; \sum_k p_k = 1\,,
\end{align}
where we define $\left| \vec{n}^k \right|^2 := \sum_{nm}
n_{nm}^*n_{nm} = \sum_{nm} |n_{nm}|^2$.

Performing the trace, we obtain (keeping notation $\rho^\dag$ formula
\eqref{ew-sepstates} becomes more evident)
\begin{equation}\label{ew-sepstates}
    \langle \rho , C \rangle \;=\; \textnormal{Tr} \rho^\dag \, C \;=\; \sum_k p_k
    \left( (d-1)a \left( 1 \,+\, \sum_{n,m} c_{nm} n_{nm}^{*k} m_{-nm}^{*k} \right) \right)\,,
\end{equation}
and using the restriction $|c_{nm}| \,\leq\, 1 \;\; \forall \, n,m $ we have
\begin{equation}
    \left| \sum_{n,m} c_{nm} n_{nm}^{*k} m_{-nm}^{*k} \right|
    \;\leq\; \sum_{n,m} |n_{nm}^k| |m_{-nm}^k| \;\leq\; 1 \,,
\end{equation}
and since the convex sum of positive terms stays positive we get $\langle \rho , C
\rangle \;\geq\; 0 \;\; \forall \rho \in S \,. \ \Box$\\

Since the operator $\tilde{C}$ \eqref{ctildequtrit} is of the form \eqref{lemqutritc} we
can use Lemma~\ref{lemqutrit} to verify
\begin{equation}
    \langle \rho , \tilde{C} \rangle \;\geq\; 0 \qquad \forall \rho \in S \,.
\end{equation}
Thus $\tilde{C}$ \eqref{ctildequtrit} is indeed an entanglement witness and
$\tilde{\rho}_\beta$ is the nearest separable state $\tilde{\rho}_\beta = \rho_{0; \,
\beta}$ for the entangled states $\rho_{\alpha,\beta}^{\rm{ent}}$ in Region I.\\

For the HS measure of the entangled two--parameter two--qutrit states \eqref{rhoqutrit}
we find
\begin{equation} \label{hsqutritreal}
    D^I(\rho_{\alpha, \beta}^{\rm{ent}}) \;=\; \| \rho_{0; \, \beta} -
    \rho_{\alpha,\beta}^{\rm{ent}} \| \;=\; \frac{2\sqrt{2}}{3}
    \left( \alpha - \frac{1}{4} - \frac{1}{8} \beta \right) \,.
\end{equation}

\subsubsection{Region II}

In Region II of Fig.~\ref{figqutrit}, the entangled two--parameter two--qutrit states are
constrained by
\begin{equation}\label{param-qutrit-regionII}
    \alpha \;<\; \frac{5}{4}\beta - \frac{1}{2}, \quad\; \alpha \;\geq\;
    \frac{1}{8}\beta - \frac{1}{8}, \quad\; \alpha \;\leq\; -\beta + 1 \,.
\end{equation}
The points that have minimal Euclidean distance to the points
$(\alpha, \beta)$ located in this region are
\begin{equation}
\left(
    \begin{array}{c}
    \tilde{\alpha} \\
    \tilde{\beta}
    \end{array}
\right) = \left(
    \begin{array}{c}
    1/24 \,\left( -2 + 20\alpha + 5\beta \right) \\
    1/6 \,\left( 2 + 4\alpha + \beta \right)
    \end{array}
\right) \,,
\end{equation}
and correspond to the states $\tilde{\rho}_{\alpha, \beta}$. The quantities needed for
calculating $\tilde{C}$ are
\begin{eqnarray}
    \tilde{\rho}_{\alpha, \beta} \,-\, \rho_{\alpha, \beta}^{\rm{ent}} &\;=\;&
    - \frac{1}{72} \left( 4\alpha + 2 - 5\beta \right) \left( U_1 - U_2 \right) \,,\\
    \| \tilde{\rho}_{\alpha, \beta} - \rho_{\alpha, \beta}^{\rm{ent}} \| &\;=\;&
    \frac{1}{6\sqrt{2}} \left( -4\alpha - 2 + 5\beta \right) \,,\\
    \langle \tilde{\rho}_{\alpha, \beta} , \tilde{\rho}_{\alpha, \beta}^{\rm{ent}} -
    \rho_{\alpha, \beta} \rangle &\;=\;& \frac{1}{36} \left( 4\alpha + 2 - 5\beta \right) \,,
\end{eqnarray}
so that operator $\tilde{C}$ is expressed by
\begin{equation} \label{ctildequtrit2}
    \tilde{C} \;=\; \frac{1}{6\sqrt{2}} \,\left( 2\,\mathbbm{1} \,+\, U_1 \,-\, U_2 \right) \,.
\end{equation}
The check of the first of conditions \eqref{defentwit} for an entanglement witness gives,
unsurprisingly,
\begin{equation}
    \langle \rho_{\alpha, \beta}^{\rm{ent}} , \tilde{C} \rangle \;=\;
    \frac{1}{6\sqrt{2}} \,\left( 4\alpha + 2 - 5\beta \right) \,<\, 0 \,,
\end{equation}
since $4\alpha < 5\beta -2\,$, Eq.~\eqref{param-qutrit-regionII}. For the second test we
use the fact that operator $\tilde{C}$ \eqref{ctildequtrit2} is of the form
\eqref{lemqutritc} and thus, according to Lemma~\ref{lemqutrit}, we obtain
\begin{equation}
    \langle \rho , \tilde{C} \rangle \;\geq\; 0 \qquad \forall \rho \in S \,.
\end{equation}
Therefore, $\tilde{C}$ \eqref{ctildequtrit2} is indeed an entanglement witness and the
states $\tilde{\rho}_{\alpha, \beta}$ are the nearest separable ones
$\tilde{\rho}_{\alpha, \beta} = \rho_{0; \, \alpha, \beta}$ to the entangled
two--parameter states \eqref{rhoqutrit} of Region II.

Finally, for the HS measure of these states, we find
\begin{equation}
    D^{II}(\rho_{\alpha,\beta}^{\rm{ent}}) \;=\; \| \rho_{0; \, \alpha, \beta} \,-\,
    \rho_{\alpha,\beta}^{\rm{ent}} \| \;=\; \frac{1}{6\sqrt{2}} \,
    \left( -4\alpha - 2 + 5\beta \right) \,.\\
\end{equation}

Another way to arrive at the nearest separable states for the two--parameter states is to
calculate the nearest PPT states with the method of Ref.~\cite{verstraete02} first and
then check if the gained states are separable. If we do so we obtain for the nearest PPT
states the states $\tilde{\rho}_\beta$ and $\tilde{\rho}_{\alpha, \beta}$ we found with
our ``guess method'', we know from Ref.~\cite{baumgartner06} these states are separable
and therefore they have to be the nearest separable states.

\subsection{Three--parameter entangled states and bound entanglement --- qutrits}
\label{secqutrits-be}

\subsubsection{Detecting bound entanglement with geometric entanglement witnesses}
\label{qutrits-be}

As we already mentioned, the PPT-criterion \cite{peres96, horodecki96} is a necessary
criterion for separability (and sufficient for $2 \times 2$ or $2 \times 3$ dimensional
Hilbert spaces). A separable state has to stay positive semidefinite under partial
transposition. Thus, if a density matrix becomes indefinite under partial transposition,
i.e. one or more eigenvalues are negative, it has to be entangled and we call it a
\emph{NPT entangled state}. But there exist entangled states that remain positive
semidefinite -- \emph{PPT entangled states} -- these are called \emph{bound entangled
states}, since they cannot be distilled to a maximally entangled state
\cite{horodecki97a, horodecki98}.

Let us consider states on a $d_1 \times d_2$ dimensional Hilbert space, $D:=d_1 d_2$. The
set of all PPT states $P$ is convex and compact and contains the set of separable states.
Thus in Eq.~\eqref{defhs} the nearest separable state $\rho_0$ can be replaced by the
nearest PPT state $\tau_0$ for which the minimal distance to the set of PPT states is
attained,
\begin{equation}
    \min_{\tau \in P} \| \tau - \rho \| \;=\; \| \tau_0 - \rho \| \,.
\end{equation}
If $\rho$ is a NPT entangled state $\rho_{\rm{NPT}}$ and $\tau_0$ the nearest PPT state,
then the operator
\begin{equation} \label{PPToperator}
A_{\rm{PPT}} \;:=\; \tau_0 \,-\, \rho_{\rm{NPT}} \,-\, \langle
\tau_0, \tau_0 - \rho_{\rm{NPT}} \rangle \,\mathbbm{1}_{\rm{D}}
\end{equation}
defines a tangent hyperplane to the set $P$ for the same geometric reasons as operator \eqref{entwitmaxviolation} and has to be an entanglement witness since $P \supset S$ (for convenience we do not normalize \eqref{PPToperator} since it does not affect the following
calculations). The nearest PPT state $\tau_0$ can be found using the method provided in Ref.~\cite{verstraete02}. In principle, the entanglement of $\rho_{\rm{NPT}}$ can be measured in experiments that should verify Tr$A_{\rm{PPT}} \,\rho_{\rm{NPT}} < 0$. If the
state $\tau_0$ is separable, it has to be the nearest separable state $\rho_0$ since the operator \eqref{PPToperator} defines a tangent hyperplane to the set of separable states. Therefore, in this case, $A_{\rm{PPT}}$ is an optimal entanglement witness,
$A_{\rm{PPT}} = A_{\rm{opt}}$, and the HS measure of entanglement can be readily obtained. If $\tau_0$ is not separable, that is PPT and entangled, it has to be a bound entangled state.

Unfortunately, it is not trivial to check if the state $\tau_0$ is separable or not. As
already mentioned, it is hard to find evidence of separability, but it might be easier to
reveal bound entanglement, not only for the state $\tau_0$ but for a whole family of
states. A method to detect bound entangled states we are going to present now.

\begin{figure}
  \includegraphics[width=0.40\textwidth]{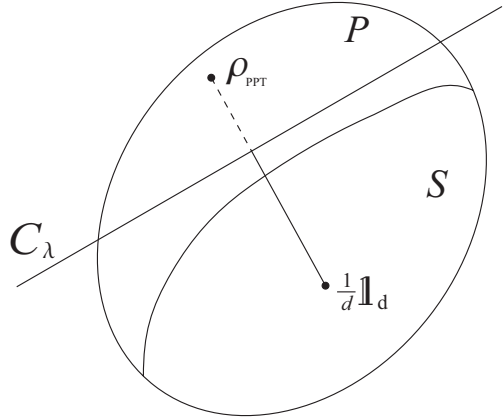}\\
  \caption{Sketch of the presented method to detect bound entanglement with the geometric
    entanglement witness $C_\lambda$. The dashed line indicates the detected bound entangled
    states $\rho_\lambda$ ($\lambda_{\rm{min}} < \lambda \leq 1$), states $\rho_\lambda$ with
    $0 < \lambda \leq \lambda_{\rm{min}}$ can be separable or bound entangled
    (straight line).}\label{figgeneral}
\end{figure}

Consider any PPT state $\rho_{\rm{PPT}}$ and the family of states $\rho_\lambda$ that lie
on the line between $\rho_{\rm{PPT}}$ and the maximally mixed (and of course separable)
state $\frac{1}{D} \mathbbm{1}_{\rm{D}}$,
\begin{equation} \label{rholambdaPPT}
    \rho_\lambda \;:=\; \lambda \,\rho_{\rm{PPT}} \,+\,  \frac{(1-\lambda)}{D}
    \,\mathbbm{1}_{\rm{D}} \,.
\end{equation}
We can construct an operator $C_\lambda$ in the following way:
\begin{equation} \label{cbe}
    C_\lambda \;=\; \rho_\lambda \,-\, \rho_{\rm{PPT}} \,-\, \langle \rho_\lambda ,
    \rho_\lambda - \rho_{\rm{PPT}} \rangle \,\mathbbm{1}_{\rm{D}} \,.
\end{equation}
If we can show that for some $\lambda_{\rm{min}} < 1$ we have
$\textnormal{Tr} \,\rho\, C_{\lambda_{\rm{min}}} \geq 0$ for all
$\rho \in S$, $C_{\lambda_{\rm{min}}}$ is an entanglement witness
(due to the construction of $C_\lambda$ the condition Tr
$\rho_\lambda C_\lambda < 0$ is already satisfied) and therefore
$\rho_{\rm{PPT}}$ and all states $\rho_\lambda$ with
$\lambda_{\rm{min}} < \lambda \leq 1$ are bound entangled (see
Fig.~\ref{figgeneral}). In Ref.~\cite{bandyopadhyay08} a similar
approach is used to identify bound entangled states in the context
of the robustness of entanglement.

\subsubsection{Application of the method to detect bound entanglement}\label{secapplmethod}

Let us now introduce the following family of three--parameter two--qutrit states:
\begin{equation} \label{famstates}
    \rho_{\alpha,\beta,\gamma} \;:=\; \frac{1-\alpha -\beta -\gamma}{9}
    \,\mathbbm{1} \,+\, \alpha P_{00} \,+\, \frac{\beta}{2} \left( P_{10} +
    P_{20} \right) \,+\, \frac{\gamma}{3} \left( P_{01} +P_{11}+P_{21}
    \right)\,,
\end{equation}
where the parameters are constrained by the positivity requirement
$\rho_{\alpha,\beta,\gamma} \geq 0$,
\begin{alignat}{2} \label{rho3param-pos}
    \alpha \;&\leq\; \frac{7}{2} \beta +1 -\gamma \,, \quad
    & \alpha \;&\leq\; -\beta +1 -\gamma \,, \nonumber\\
    \alpha \;&\leq\; -\beta +1 +2\gamma \,, \quad
    & \alpha \;&\geq\; \frac{\beta}{8} - \frac{1}{8} + \frac{1}{8}\gamma \,.
\end{alignat}

States \eqref{famstates} lie again in the magic simplex and for
$\gamma = 0$ we come back to the states \eqref{rhoqutrit} considered
before. However, for $\gamma \neq 0$ it is not trivial to find the
nearest separable states since the PPT states do not necessarily
coincide with the separable states. But we can use our geometric
entanglement witness \eqref{cbe} to detect bound entanglement (see
Refs.~\cite{bertlmann08, bertlmann08b}).

We start with the following one--parameter family of two--qutrit states that was
introduced in Ref.~\cite{horodecki99c}:
\begin{equation} \label{horstates}
    \rho_b \;=\; \frac{2}{7} \left| \phi_+^3 \right\rangle \left\langle
    \phi_+^3 \right| \,+\, \frac{b}{7} \, \sigma_+ \,+\, \frac{5 - b}{7} \,
    \sigma_- \,, \qquad 0 \leq b \leq 5 \,,
\end{equation}
where
\begin{align}
    \sigma_+ \;:=\; \ & \frac{1}{3} \left( \left| 0 1 \right\rangle
        \left\langle 0 1 \right| \,+\, \left| 1 2 \right\rangle \left\langle
        1 2 \right| \,+\, \left| 2 0 \right\rangle \left\langle 2 0 \right|\,
        \right) \,, \\
    \sigma_- \;:=\; \ & \frac{1}{3} \left( \left| 1 0\right\rangle
        \left\langle 1 0 \right| \,+\, \left| 2 1 \right\rangle \left\langle
        2 1 \right| \,+\, \left| 0 2 \right\rangle \left\langle 0 2 \right|\,
    \right) \,.
\end{align}
Let us call this family of states \emph{Horodecki states}. Interestingly, the states
\eqref{horstates} are part of the three--parameter family \eqref{famstates}, namely
\begin{equation} \label{horstatessimplex}
    \rho_b \;\equiv\; \rho_{\alpha, \beta, \gamma} \qquad\mbox{with}\quad \alpha =
    \frac{6-b}{21}, \;\beta = -\frac{2b}{21}, \;\gamma = \frac{5-2b}{7} \,,
\end{equation}
and thus lie in the magic simplex. Testing the partial transposition, we find that the Horodecki states \eqref{horstates} split the states in the following way: for $0 \leq b < 1$ they are NPT, for $1 \leq b \leq 4\,$ PPT and for $4 < b \leq 5\,$ NPT. In Ref.~\cite{horodecki99c}, it is shown that the states are separable for $2 \leq b \leq 3$ and bound entangled for $3 < b \leq 4\,$. In our case, it is more convenient to use $\gamma$ as the parameter of the
Horodecki states. Using Eq.~\eqref{horstatessimplex} we express $b$
in terms of $\gamma$ and obtain
\begin{equation} \label{horstatessimplexgamma}
    \rho_b \;\equiv\; \rho_{\alpha, \beta, \gamma} \qquad\mbox{with}\quad \alpha =
    \frac{1+\gamma}{6}, \beta= \frac{-5+7\gamma}{21}, \gamma \,.
\end{equation}

\begin{figure}
  \includegraphics[width=0.6\textwidth]{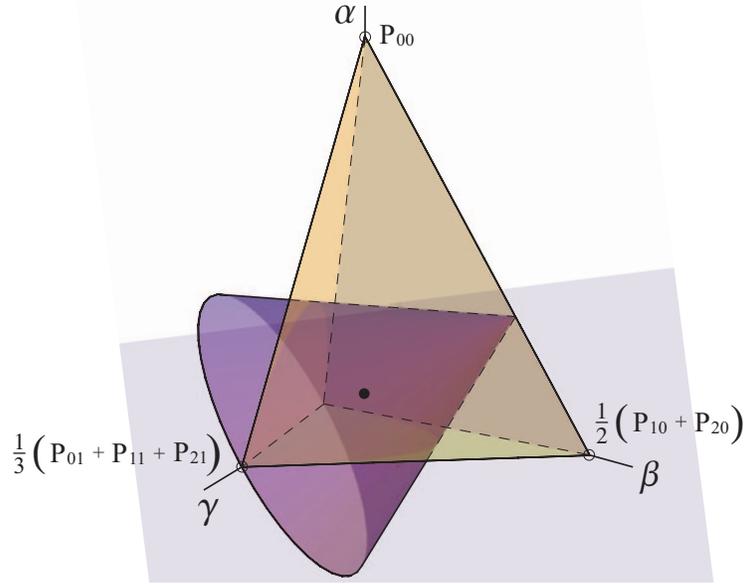}\\
  \caption{The states $\rho_{\alpha, \beta, \gamma}$ represent a pyramid with a triangular
    base due to the positivity constraints \eqref{rho3param-pos} and the PPT points with
    constraints \eqref{rho3paramPPT} form a cone which overlaps with the pyramid. The
    bound entangled and separable states lie in the intersection region. The black dot
    indicates the maximally mixed state and the origin of the non--orthogonal coordinate
    axes.}
  \label{simplex3d_fit}
\end{figure}

The geometry of the three--parameter family of states $\rho_{\alpha,\beta,\gamma}$ as
part of the magic simplex we show in Fig.~\ref{simplex3d_fit},
in particular, the states $\rho_{\alpha, \beta, \gamma}$ with positivity requirement
\eqref{rho3param-pos} and the PPT states which are constrained by
\begin{eqnarray} \label{rho3paramPPT}
    \alpha &\;\leq\;& - \beta - \frac{1}{2} + \frac{1}{2}\gamma \,, \nonumber\\
    \alpha &\;\leq\;& \frac{1}{16} \left( -2 + 11\beta + 3 \sqrt{\Delta}\right)\,, \;\quad
    \alpha \;\leq\; \frac{1}{16} \left( -2 + 11\beta - 3 \sqrt{\Delta}\right) \,,
\end{eqnarray}
where $\Delta = 4 + 9 \beta^2 + 4\gamma - 7\gamma^2 - 6\beta (2 + \gamma)$. The states $\rho_{\alpha, \beta, \gamma}$ form due to the positivity constraints \eqref{rho3param-pos} a pyramid with triangular base and the PPT points due to the constraints \eqref{rho3paramPPT} a cone. Both objects are quite symmetric and overlap with each other in a way shown in Fig.~\ref{simplex3d_fit}. In the intersection region lie the bound entangled and separable
states. \\

\noindent\emph{Application of the method.} We now want to apply the method to detect bound entanglement of the three--parameter family \eqref{famstates}. The idea is to choose PPT starting points on the boundary plane $\alpha = \frac{7}{2} \beta +1 - \gamma$ of the
positivity pyramid, on the Horodecki line and in a region close to this line, and shift the operators $C_\lambda$ along the parameterized lines that connect the starting points with the
maximally mixed states. If we can show that $C_{\lambda}$ is an entanglement witness until a certain $\lambda_{min}$, all states $\rho_\lambda$ \eqref{rholambdaPPT} with $1 \leq \lambda <
\lambda_{min}$ are PPT entangled.

We parameterize our ``starting states'' on the boundary plane by
\begin{equation}
    \rho_{\rm{plane}} \;\equiv\; \rho_{\alpha, \beta, \gamma} \quad \mbox{with} \quad
    \left(\alpha = \frac{1+\gamma+\epsilon}{6} \,, \beta=
    \frac{-5+7\gamma+\epsilon}{21} \,,\gamma \right), \quad \epsilon \in
    \mathbbm{R}\,,
\end{equation}
where we introduced an additional parameter $\epsilon$ to account
for the deviation from the line within the boundary plane.

Depending on $\gamma$ and $\epsilon$ the operator $C_{\gamma,
\epsilon, \lambda}$ \eqref{cbe} has the following form:
\begin{align} \label{cbesimplexvol}
    &C_{\gamma, \lambda} \,\;=\;\, \rho_\lambda \,-\, \rho_{pl} \,-\,
    \langle \rho_\lambda , \rho_\lambda - \rho_{pl} \rangle \,\mathbbm{1}
    \,\;=\;\, a \,(2 \,\mathbbm{1} \,+\, c_1 U_1 \,+\, c_2 U^I_2 \,+\, c^{\star}_2 U^{II}_2)
    \,, \nonumber\\
    &\mbox{with}\quad a \;=\; \frac{d}{36} \,\lambda (1-\lambda)\,, \quad
    d \;=\; 1 + 3\gamma ^2 + 3\epsilon (2+\epsilon )/7 \,, \nonumber\\
    &c_1 \;=\; -\frac{4 (2+\epsilon )}{7d\lambda} \,, \quad
    c_2 \;=\; \frac{2(1 - 7 \sqrt{3} \gamma \,i - 3\epsilon )}{7d\lambda}
    \,.
\end{align}
The operators $U_1, U^I_2, U^{II}_2$ are defined by
Eq.~\eqref{defu1u2} and the family of states $\rho_\lambda$ by
\begin{equation} \label{rholambdavol}
    \rho_\lambda \;=\; \lambda \,\rho_{\rm{plane}} \,+\, \frac{1-\lambda}{9} \,
    \mathbbm{1} \,.
\end{equation}
We want to find the minimal $\lambda$, denoted by $\lambda_{\rm{min}}$, depending on the
parameters $\gamma$ and $\epsilon\,$, such that all states on the line \eqref{rholambdavol}
are bound entangled for $\lambda_{\rm{min}} < \lambda \leq 1$. To accomplish this, we
define the functions
\begin{equation}
    g_1(\gamma, \epsilon, \lambda) \;:=\; |c_1| \quad \mbox{and} \quad g_2(\gamma,
    \epsilon, \lambda) \;:=\; |c_2| \,,
\end{equation}
then $\lambda_{\rm{min}}$ is attained at $\max \{ g_1(\gamma, \epsilon, \lambda),
g_2(\gamma, \epsilon, \lambda) \} =1$ (recall Lemma~\ref{lemqutrit}). Bound entanglement
can be found in a region where $\lambda_{\rm{min}} < 1$ and the starting points of the
lines \eqref{rholambdavol} are PPT states. That means, $\epsilon$ and $\gamma$ are chosen
such that the starting points are PPT states and the corresponding line allows a
$\lambda_{\rm{min}} < 1$. The parameter $\epsilon$ is bounded by $-1/4 < \epsilon <
1/3\,$, where the lower bound is reached for $\lambda_{\rm{min}} \rightarrow 1$ at
$|\gamma| = 1/4$ and the upper bound arises from the boundary of the PPT states at
$\gamma = 0$. For every $\epsilon$ in this interval, we have an interval of $|\gamma|$
where bound entangled states are located.
\begin{figure}
  \includegraphics[width=0.5\textwidth]{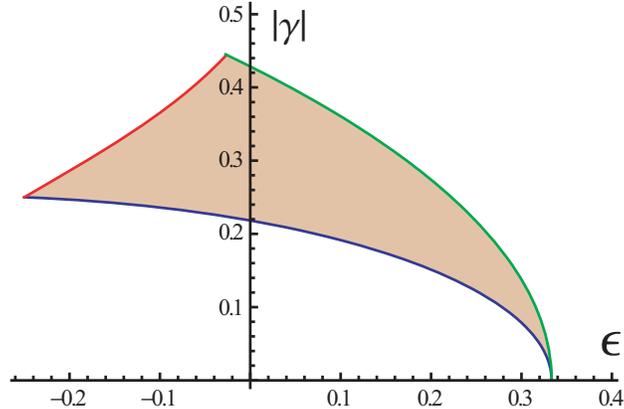}\\
  \caption{Plot of the parameter range $\gamma$ versus $\epsilon$ of
    starting points of the lines \eqref{rholambdavol} such that bound
    entangled states can be detected.} \label{lambdaplot3d}
\end{figure}

\begin{figure}
  \includegraphics[width=0.7\textwidth]{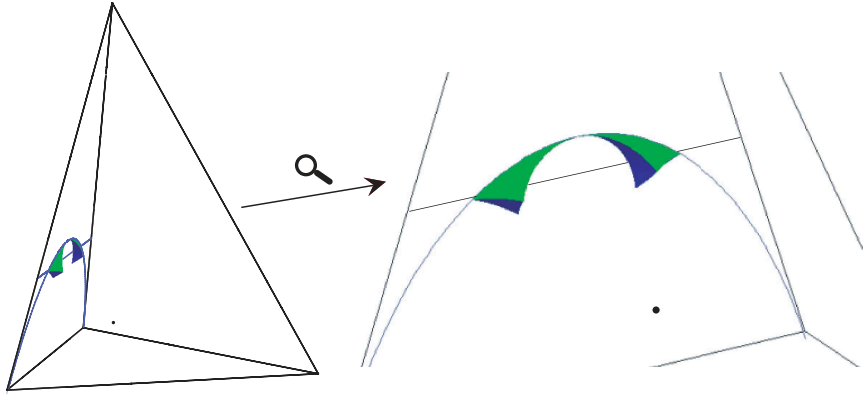}\\
  \caption{Regions of detected bound entangled states within the pyramid represented
    by the three--parameter states $\rho_{\alpha ,\beta ,\gamma}$ \eqref{famstates}.
    The dot represents the maximally mixed state, the Horodecki states are represented by the
    line through the boundary plane from which the regions of bound entanglement emerge.
    In the magnified picture on the right
    hand side the viewpoint is altered a bit to gain a better visibility.}
    \label{volumes}
\end{figure}

More precisely, in the interval for the $\epsilon$ parameter $-1/4 < \epsilon <
\epsilon_0\,$ with $\epsilon_0 = \left(8 \,-\, 7\,(2/(-5+\sqrt{29}))^{1/3} \,+\,
7\,(2/(-5+\sqrt{29}))^{-1/3}\right)/\,3 \;\simeq\; -\,0.03\,$ the parameter $\gamma$ is
confined by $\sqrt{1 - 2\epsilon + 3 \epsilon ^2}\,/\sqrt{21} < |\gamma| < \sqrt{7 -
6\epsilon - 3\epsilon ^2 - 2\,(1 - 48\epsilon - 12\epsilon ^2)^{1/2}}\,/\sqrt{21}\,$, under
the constraint $\lambda_{\rm{min}} < 1$. For the remaining interval $\epsilon_0 <
\epsilon < 1/3\,$, we get the bounds $\sqrt{1 - 2\epsilon + 3 \epsilon ^2}\,/\sqrt{21} <
|\gamma| < \sqrt{9 - 26\epsilon - 3\epsilon ^2}\,/\,7\,$, where the lower bound is again
constrained by $\lambda_{\rm{min}} < 1$ and the upper one by the PPT condition. A plot of
the allowed values of $\epsilon$ and $\gamma$ for the starting points on the boundary
plane is depicted in Fig.~\ref{lambdaplot3d}. We have equality of the coefficient
functions $g_1 = g_2$ for $|\gamma| = \sqrt{15 + 22\epsilon - 5\epsilon ^2}\,/\,7
\sqrt{3} =: \gamma_0\,$, $g_1 > g_2$ for $|\gamma| < \gamma_0$ and $g_1 < g_2$ for
$|\gamma| > \gamma_0$, where $|\gamma|$ is always restricted to the allowed range
described above. As mentioned, $\lambda_{\rm{min}}$ is gained from the condition $\max \{
g_1(\gamma, \epsilon, \lambda), g_2(\gamma, \epsilon, \lambda) \} = 1$ for particular
values of $\gamma$ and $\epsilon$. The total minimum $\lambda_{\rm{min}}^{\rm{tot}}$ is
finally reached at
\begin{equation}\label{lambda-min-vol}
    \lambda_{\rm{min}}^{\rm{tot}} \;=\; \frac{1}{8} \,(3 + \sqrt{13}) \;\simeq\;
    0.826 \,,
\end{equation}
which is significantly below the value $1$ so that the resulting volume of bound entangled states is remarkably large. The total minimum \eqref{lambda-min-vol} is attained at $\epsilon = (7\sqrt{13}-25)\,/\,2 \;\simeq\; 0.12$ and $|\gamma| \;\simeq\; 0.35\,$. The whole line
of states $\rho_\lambda$ \eqref{rholambdavol} within the interval
$\lambda_{\rm{min}}^{\rm{tot}} < \lambda \leq 1$ is found to be bound entangled. The volume of the detected bound entangled states we have visualized in Fig.~\ref{volumes}. For $\gamma = 0$ no bound entanglement occurs, as discussed in Sec.~\ref{secqutrits}, which is represented in Fig.~\ref{volumes} at the meeting point of the two bound entangled regions.

\subsubsection{More bound entangled states and the shape of
    separable states} \label{secmorebe}

In the last section, we gave a strict application of our method to
detect bound entangled states, where we recognized bound entangled
states on the parameterized lines $\rho_\lambda$
\eqref{rholambdaPPT} only. The involved entanglement witnesses,
however, are able to detect the entanglement of \emph{all} states on
one side of the corresponding plane, not only on the lines.
Therefore even larger regions of bound entanglement can be
identified for the three--parameter family \eqref{famstates}, which
is described in detail in Ref.~\cite{bertlmann08b}.

In this subsection, we want to show that for the three--parameter
family \eqref{famstates} our method detects the same bound entangled
states as the realignment criterion does, and furthermore allows for
a construction of the shape of the separable states of the
three--parameter family, so that we are able to fully determine the
entanglement properties of this family of states.

The realignment criterion is a necessary criterion of separability
and says that for any separable state the sum of the singular values
$s_i$ of a \emph{realigned} density matrix $\sigma_R$ has to be
smaller than or equal to one,
\begin{equation} \label{realign}
    \sum_i s_i \;=\; \textnormal{Tr} \sqrt{\sigma^\dag_R \sigma_R}\;\leq\; 1 \,,
\end{equation}
where $\left( \rho_{ij,kl} \right)_R := \rho_{ik,jl}$ (for details
see Refs.~\cite{rudolph00, rudolph02, rudolph03, chen03}). States
that violate the criterion have to be entangled, states that satisfy
it can be entangled or separable.

In our case of the three--parameter family \eqref{famstates} we
obtain the constraints
\begin{align}
    \alpha \;\leq\; & \ \frac{1}{16} \left(6+11 \beta -\gamma
        - \Delta_1 \right) \label{realignconstraint} \\
    \alpha \;\leq\; & \ \frac{1}{16} \left(6+11 \beta -\gamma
        + \Delta_1 \right) \\
    \alpha \;\geq\; & \ \frac{1}{16} \left(-6+11 \beta -\gamma
        - \Delta_2 \right) \\
    \alpha \;\geq\; & \ \frac{1}{16} \left(-6+11 \beta -\gamma
        + \Delta_2 \right)
\end{align}
from the realignment criterion, where
\begin{align}
    \Delta_1 \;:=\; & \ \sqrt{4+36 \beta +81 \beta ^2-12 \gamma -54 \beta \gamma
        +33 \gamma ^2} \qquad \mbox{and} \nonumber\\
    \Delta_2 \;:=\; & \ \sqrt{4-36 \beta +81 \beta ^2+12
        \gamma -54 \beta  \gamma +33 \gamma ^2}.
\end{align}
Only constraint \eqref{realignconstraint} is violated by some PPT
states, which thus have to be bound entangled. The PPT entangled
states exposed by the realignment criterion are therefore
concentrated in the region confined by the constraints
\begin{equation} \label{famstatesbe}
\alpha \;\leq\; \frac{7}{2} \beta +1 -\gamma, \;\ \alpha \;\leq\; \frac{1}{16}
\left( -2 + 11\beta + 3 \sqrt{\Delta}\right), \;\ \alpha \;\geq\; \
\frac{1}{16} \left(6+11 \beta -\gamma - \Delta_1 \right) \,.
\end{equation}
All PPT entangled states of Eq.~\eqref{famstatesbe} can also be
detected using Lemma~\ref{lemqutrit}. To see this, we construct
tangent planes onto the surface of the function
\begin{equation} \label{realignfunction}
\alpha \;=\; \ \frac{1}{16} \left(6+11 \beta -\gamma - \Delta_1 \right)
\end{equation}
from the realignment criterion \eqref{realignconstraint}, where we
use orthogonal coordinates. In this way, we can assign geometric
operators to the tangential planes by choosing points $\vec{a}$
inside the planes and points $\vec{b}$ outside the planes such that
$\vec{a}- \vec{b}$ is orthogonal to the planes. Since the Euclidean
geometry of our picture is isomorph to the Hilbert-Schmidt geometry,
the points $\vec{a}$ and $\vec{b}$ correspond to states $\rho_a$ and
$\rho_b$ and we can construct an operator accordingly,
\begin{equation} \label{geomopre}
C_{\rm{re}} \;=\; \rho_a - \rho_b \;-\; \langle \rho_a, \rho_a - \rho_b
\rangle \mathbbm{1}_9 \,.
\end{equation}

\begin{figure}
  \includegraphics[width=0.4\textwidth]{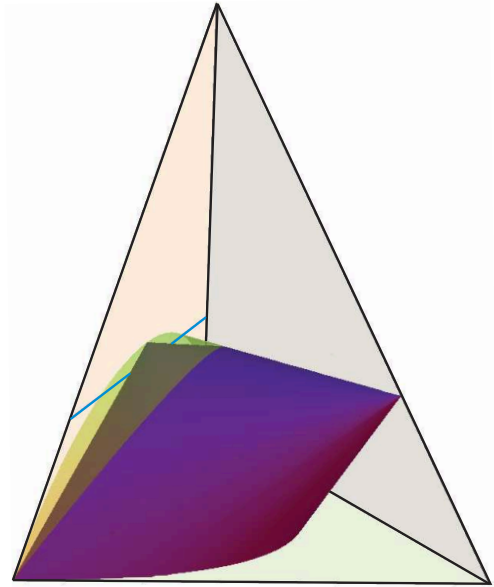}\\
  \caption{Illustration of the PPT and realignment criteria. The
  familiar PPT cone (here pictured only inside the positivity
  pyramid) is intersected by a cone formed by the realignment
  constraint \eqref{realignconstraint} and reveals a region of bound
  entangled states (translucent yellow region) that can also be
  detected by Lemma \ref{lemqutrit}.} \label{fullent}
\end{figure}

Decomposed into the Weyl operator basis, the operators
\eqref{geomopre} that correspond to tangent planes in points
$(\alpha_{\rm t}, \beta_{\rm t}, \gamma_{\rm t})$ -- where
$\alpha_{\rm t}$ is a function of $\beta_{\rm t}$ and $\gamma_{\rm
t}$, given by the realignment function \eqref{realignfunction} --
are
\begin{align} \label{geomopreweyl}
    C_{\rm{re}} \;=\; & \ a \,(2 \,\mathbbm{1} - U_1 + c \, U^I_2 \,+\, c^* U^{II}_2 )
        \,, \quad \mbox{with} \nonumber\\
    &a \;=\; \frac{1}{36} \left(-2-9 \beta_{\rm t}+3 \gamma_{\rm t}+3
        \Delta_c \right), \nonumber\\
    &c \;=\; \frac{9 \gamma_{\rm t}^2+(-2-9 \beta_{\rm t}+3 \gamma_{\rm t})
        \Delta_c + \sqrt{3} \gamma_{\rm t} \left(2+9 \beta_{\rm t}-3 \gamma_{\rm t}+3
        \Delta_c \right) i}{(2+9 \beta_{\rm t})^2-6 (2+9 \beta_{\rm t}) \gamma_{\rm t}
        +36 \gamma_{\rm t}^2} \,, \nonumber\\
    &\Delta_c \;:=\; \sqrt{4+36 +81 \beta_{\rm t}^2-12 \gamma_{\rm t}-54
        \beta_{\rm t} \gamma_{\rm t}+33 \gamma_{\rm t}^2} \;.
\end{align}
The absolute values of the coefficients $c$ and $c^*$ in
Eq.~\eqref{geomopreweyl} are $1$, $|c| = |c^*| = 1$, and therefore,
according to Lemma~\ref{lemqutrit}, the operators $C_{\rm{re}}$ are
entanglement witnesses that detect the entanglement of all states
``above'' the corresponding planes, thus also the bound entangled
states in the region of Eq.~\eqref{famstatesbe}. The detected bound
entangled region is depicted in Fig.~\ref{fullent}.
Naturally, now the question arises if all the states $\rho_{\alpha,
\beta, \gamma}$ \eqref{famstates} that satisfy both the PPT and the
realignment criterion are separable. This cannot be seen using the
two criteria alone, since they are not sufficient criteria for
separability. But we can apply a method of shifting operators along
parameterized lines in the other direction: by constructing a kernel
polytope of necessarily separable states and assigning operators to
the boundary planes of this polytope, we can shift those operators
outside until they become entanglement witnesses, which can be shown
numerically (unfortunately Lemma~\ref{lemqutrit} does not help
here). In this way, one can step by step reconstruct the shape of the
separable states, which indeed is the set given by the states that
satisfy the PPT and the realignment criterion (see dark shape of the
two intersecting cones in Fig.~\ref{fullent}); for details see
Ref.~\cite{krammer08}. Hence, we see that the presented method of ``shifting''
operators along parameterized lines does not only help to detect
bound entanglement, but also to construct the shape of the separable
states.\\

\section{Conclusion}
\label{conclusion}

We discuss the geometric aspects of entanglement for density matrices within a simplex
formed by Bell states. We use entanglement witnesses in order to quantify entanglement
and detect in case of qutrits bound entangled states in specific instances.

We demonstrate the geometry of separability and entanglement in case of qubits by
choosing so-called two--parameter states, Eq.~\eqref{rhoqubit}, i.e., planes in the
simplex, a tetrahedron (see Fig.~\ref{figtetrahedron}). These states reflect already the
underlying geometry of the Hilbert space and they are chosen with regard to the
description of qutrit states, a generalization into higher dimensions. To a given
entangled state we determine the nearest separable state, calculate the corresponding
entanglement witness and the Hilbert--Schmidt measure in the relevant Regions I and II
(see Fig.~\ref{figqubit}).

In case of qutrits it is quite illustrative to demonstrate the geometry of separability
and entanglement in terms of two--parameter states \eqref{rhoqutrit}. These states
represent a plane in the eight--dimensional simplex formed by the nine Bell states, the magic
simplex, and are easy to construct within the Weyl operator basis. Due to the high
symmetry of this simplex we may restrict ourselves to a certain mixture of Bell states,
Eq.~\eqref{rhoqutrit}, which lie on a line in a phase space description. This line
exhibits the same geometry as other lines. Within the Weyl operator basis it is quite
easy to find the Bloch vector form \eqref{rhoqutritweyl} of the two--parameter states. It
enables us to find in regions I and II (see Fig.~\ref{figqutrit}) the nearest separable
state to a given entangled state and the corresponding entanglement witness. The easy
calculation of the Hilbert--Schmidt measure of entanglement is a great advantage and its
result of high interest. Other entanglement measures, like the entanglement of formation,
are much harder to be calculated in this higher dimensional case.

We present a method to find analytically bound entangled states by using entanglement witnesses. These witnesses are constructed geometrically, Eq.~\eqref{cbe}, in quite the same way as for the detection of the nearest separable state, Eq.~\eqref{entwitmaxviolation}. We show that the Horodecki states \eqref{horstates} can be described by the family of three--parameter
states \eqref{famstates} and are therefore part of the magic simplex. Geometrically, they form a line on the boundary of the pyramid represented by the three--parameter states \eqref{famstates} (see Figs.~\ref{volumes} and \ref{fullent}). We apply our method to find regions of bound entangled states within the pyramid of states \eqref{famstates} (see Fig.~\ref{volumes}). Even when restricting ourselves strictly to consider the detected bound entangled states
on the parameterized lines \eqref{rholambdaPPT} only, we find large regions of bound entanglement. Employing the realignment criterion of separability, we can reveal larger regions of bound entanglement that are also detected by Lemma~\ref{lemqutrit}. Finally, we can
apply our method of shifting operators along parameterized lines together with numerical calculations to show that there do not exist more PPT entangled states for the three--parameter family. Hence, the shape of the separable states can be constructed.

When decomposing density matrices into operator bases the Weyl operator basis is the optimal one for all our calculations. The reason is that entangled states -- the maximally entangled Bell
states -- are in fact easily constructed by unitary operators \`a la Weyl, see Eq.~\eqref{projbell}.

\begin{acknowledgments}

We would like to thank Beatrix Hiesmayr and Heide Narnhofer for helpful discussions. This
research has been financially supported by the FWF project CoQuS No W1210-N16 of the
Austrian Science Foundation and by the F140-N Research Grant of the University of Vienna.

\end{acknowledgments}

\bibliography{references}

\begin{thebibliography}{37}
\expandafter\ifx\csname natexlab\endcsname\relax\def\natexlab#1{#1}\fi
\expandafter\ifx\csname bibnamefont\endcsname\relax
  \def\bibnamefont#1{#1}\fi
\expandafter\ifx\csname bibfnamefont\endcsname\relax
  \def\bibfnamefont#1{#1}\fi
\expandafter\ifx\csname citenamefont\endcsname\relax
  \def\citenamefont#1{#1}\fi
\expandafter\ifx\csname url\endcsname\relax
  \def\url#1{\texttt{#1}}\fi
\expandafter\ifx\csname urlprefix\endcsname\relax\def\urlprefix{URL }\fi
\providecommand{\bibinfo}[2]{#2}
\providecommand{\eprint}[2][]{\url{#2}}

\bibitem[{\citenamefont{Bertlmann and Zeilinger}(2002)}]{bertlmann02a}
\bibinfo{editor}{\bibfnamefont{R.~A.} \bibnamefont{Bertlmann}}
  \bibnamefont{and}
  \bibinfo{editor}{\bibfnamefont{A.}~\bibnamefont{Zeilinger}}, eds.,
  \emph{\bibinfo{title}{{Quantum [un]speakables, from Bell to quantum
  information}}} (\bibinfo{publisher}{Springer, Berlin Heidelberg New York},
  \bibinfo{year}{2002}).

\bibitem[{\citenamefont{Bouwmeester et~al.}(2000)\citenamefont{Bouwmeester,
  Ekert, and Zeilinger}}]{bouwmeester00}
\bibinfo{editor}{\bibfnamefont{D.}~\bibnamefont{Bouwmeester}},
  \bibinfo{editor}{\bibfnamefont{A.}~\bibnamefont{Ekert}}, \bibnamefont{and}
  \bibinfo{editor}{\bibfnamefont{A.}~\bibnamefont{Zeilinger}}, eds.,
  \emph{\bibinfo{title}{The physics of quantum information: quantum
  cryptography, quantum teleportation, quantum computation}}
  (\bibinfo{publisher}{Springer, Berlin, Heidelberg, New York},
  \bibinfo{year}{2000}).

\bibitem[{\citenamefont{Nielsen and Chuang}(2000)}]{nielsen00}
\bibinfo{author}{\bibfnamefont{M.}~\bibnamefont{Nielsen}} \bibnamefont{and}
  \bibinfo{author}{\bibfnamefont{I.}~\bibnamefont{Chuang}},
  \emph{\bibinfo{title}{Quantum Computation and Quantum Information}}
  (\bibinfo{publisher}{Cambridge University Press, Cambridge, England},
  \bibinfo{year}{2000}).

\bibitem[{\citenamefont{Peres}(1996)}]{peres96}
\bibinfo{author}{\bibfnamefont{A.}~\bibnamefont{Peres}},
  \bibinfo{journal}{Phys. Rev. Lett.} \textbf{\bibinfo{volume}{77}},
  \bibinfo{pages}{1413} (\bibinfo{year}{1996}).

\bibitem[{\citenamefont{Horodecki et~al.}(1996)\citenamefont{Horodecki,
  Horodecki, and Horodecki}}]{horodecki96}
\bibinfo{author}{\bibfnamefont{M.}~\bibnamefont{Horodecki}},
  \bibinfo{author}{\bibfnamefont{P.}~\bibnamefont{Horodecki}},
  \bibnamefont{and}
  \bibinfo{author}{\bibfnamefont{R.}~\bibnamefont{Horodecki}},
  \bibinfo{journal}{Phys. Lett. A} \textbf{\bibinfo{volume}{223}},
  \bibinfo{pages}{1} (\bibinfo{year}{1996}).

\bibitem[{\citenamefont{Horodecki et~al.}(2001)\citenamefont{Horodecki,
  Horodecki, and Horodecki}}]{horodecki01}
\bibinfo{author}{\bibfnamefont{M.}~\bibnamefont{Horodecki}},
  \bibinfo{author}{\bibfnamefont{P.}~\bibnamefont{Horodecki}},
  \bibnamefont{and}
  \bibinfo{author}{\bibfnamefont{R.}~\bibnamefont{Horodecki}}, in
  \emph{\bibinfo{booktitle}{Quantum Information}}, edited by
  \bibinfo{editor}{\bibfnamefont{G.~A.} \bibnamefont{et~al.}}
  (\bibinfo{publisher}{Springer Verlag Berlin}, \bibinfo{year}{2001}), vol.
  \bibinfo{volume}{173} of \emph{\bibinfo{series}{Springer Tracts in Modern
  Physics}}, p. \bibinfo{pages}{151}.

\bibitem[{\citenamefont{Horodecki}(1997)}]{horodecki97a}
\bibinfo{author}{\bibfnamefont{P.}~\bibnamefont{Horodecki}},
  \bibinfo{journal}{Phys. Lett. A} \textbf{\bibinfo{volume}{232}},
  \bibinfo{pages}{333} (\bibinfo{year}{1997}).

\bibitem[{\citenamefont{Horodecki et~al.}(1998)\citenamefont{Horodecki,
  Horodecki, and Horodecki}}]{horodecki98}
\bibinfo{author}{\bibfnamefont{M.}~\bibnamefont{Horodecki}},
  \bibinfo{author}{\bibfnamefont{P.}~\bibnamefont{Horodecki}},
  \bibnamefont{and}
  \bibinfo{author}{\bibfnamefont{R.}~\bibnamefont{Horodecki}},
  \bibinfo{journal}{Phys. Rev. Lett.} \textbf{\bibinfo{volume}{80}},
  \bibinfo{pages}{5239} (\bibinfo{year}{1998}).

\bibitem[{\citenamefont{Horodecki et~al.}(1999)\citenamefont{Horodecki,
  Horodecki, and Horodecki}}]{horodecki99c}
\bibinfo{author}{\bibfnamefont{P.}~\bibnamefont{Horodecki}},
  \bibinfo{author}{\bibfnamefont{M.}~\bibnamefont{Horodecki}},
  \bibnamefont{and}
  \bibinfo{author}{\bibfnamefont{R.}~\bibnamefont{Horodecki}},
  \bibinfo{journal}{Phys. Rev. Lett} \textbf{\bibinfo{volume}{82}},
  \bibinfo{pages}{1056} (\bibinfo{year}{1999}).

\bibitem[{\citenamefont{Rains}(1999)}]{rains99}
\bibinfo{author}{\bibfnamefont{E.~M.} \bibnamefont{Rains}},
  \bibinfo{journal}{Phys. Rev. A} \textbf{\bibinfo{volume}{60}},
  \bibinfo{pages}{179} (\bibinfo{year}{1999}).

\bibitem[{\citenamefont{Baumgartner et~al.}(2007)\citenamefont{Baumgartner,
  Hiesmayr, and Narnhofer}}]{baumgartner07}
\bibinfo{author}{\bibfnamefont{B.}~\bibnamefont{Baumgartner}},
  \bibinfo{author}{\bibfnamefont{B.~C.} \bibnamefont{Hiesmayr}},
  \bibnamefont{and}
  \bibinfo{author}{\bibfnamefont{H.}~\bibnamefont{Narnhofer}},
  \bibinfo{journal}{J. Phys. A: Math. Theor.} \textbf{\bibinfo{volume}{40}},
  \bibinfo{pages}{7919} (\bibinfo{year}{2007}).

\bibitem[{\citenamefont{Terhal}(2000)}]{terhal00}
\bibinfo{author}{\bibfnamefont{B.~M.} \bibnamefont{Terhal}},
  \bibinfo{journal}{Phys. Lett. A} \textbf{\bibinfo{volume}{271}},
  \bibinfo{pages}{319} (\bibinfo{year}{2000}).

\bibitem[{\citenamefont{Terhal}(2002)}]{terhal02}
\bibinfo{author}{\bibfnamefont{B.~M.} \bibnamefont{Terhal}},
  \bibinfo{journal}{Theoretical Computer Science}
  \textbf{\bibinfo{volume}{287}}, \bibinfo{pages}{313} (\bibinfo{year}{2002}).

\bibitem[{\citenamefont{Bertlmann et~al.}(2002)\citenamefont{Bertlmann,
  Narnhofer, and Thirring}}]{bertlmann02}
\bibinfo{author}{\bibfnamefont{R.~A.} \bibnamefont{Bertlmann}},
  \bibinfo{author}{\bibfnamefont{H.}~\bibnamefont{Narnhofer}},
  \bibnamefont{and} \bibinfo{author}{\bibfnamefont{W.}~\bibnamefont{Thirring}},
  \bibinfo{journal}{Phys. Rev. A} \textbf{\bibinfo{volume}{66}},
  \bibinfo{pages}{032319} (\bibinfo{year}{2002}).

\bibitem[{\citenamefont{Baumgartner et~al.}(2006)\citenamefont{Baumgartner,
  Hiesmayr, and Narnhofer}}]{baumgartner06}
\bibinfo{author}{\bibfnamefont{B.}~\bibnamefont{Baumgartner}},
  \bibinfo{author}{\bibfnamefont{B.~C.} \bibnamefont{Hiesmayr}},
  \bibnamefont{and}
  \bibinfo{author}{\bibfnamefont{H.}~\bibnamefont{Narnhofer}},
  \bibinfo{journal}{Phys. Rev. A} \textbf{\bibinfo{volume}{74}},
  \bibinfo{pages}{032327} (\bibinfo{year}{2006}).

\bibitem[{\citenamefont{Baumgartner et~al.}(2008)\citenamefont{Baumgartner,
  Hiesmayr, and Narnhofer}}]{baumgartner08}
\bibinfo{author}{\bibfnamefont{B.}~\bibnamefont{Baumgartner}},
  \bibinfo{author}{\bibfnamefont{B.~C.} \bibnamefont{Hiesmayr}},
  \bibnamefont{and}
  \bibinfo{author}{\bibfnamefont{H.}~\bibnamefont{Narnhofer}},
  \bibinfo{journal}{Phys. Lett. A} \textbf{\bibinfo{volume}{372}},
  \bibinfo{pages}{2190} (\bibinfo{year}{2008}).

\bibitem[{\citenamefont{Bertlmann and
  Krammer}(2008{\natexlab{a}})}]{bertlmann08a}
\bibinfo{author}{\bibfnamefont{R.~A.} \bibnamefont{Bertlmann}}
  \bibnamefont{and} \bibinfo{author}{\bibfnamefont{P.}~\bibnamefont{Krammer}},
  \bibinfo{journal}{J. Phys. A: Math. Theor.} \textbf{\bibinfo{volume}{41}},
  \bibinfo{pages}{235303} (\bibinfo{year}{2008}{\natexlab{a}}).

\bibitem[{\citenamefont{Narnhofer}(2006)}]{narnhofer06}
\bibinfo{author}{\bibfnamefont{H.}~\bibnamefont{Narnhofer}},
  \bibinfo{journal}{J. Phys. A: Math. Gen.} \textbf{\bibinfo{volume}{39}},
  \bibinfo{pages}{7051} (\bibinfo{year}{2006}).

\bibitem[{\citenamefont{Bennett et~al.}(1993)\citenamefont{Bennett, Brassard,
  Cr{\'e}peau, Jozsa, Peres, and Wootters}}]{bennett93}
\bibinfo{author}{\bibfnamefont{C.~H.} \bibnamefont{Bennett}},
  \bibinfo{author}{\bibfnamefont{G.}~\bibnamefont{Brassard}},
  \bibinfo{author}{\bibfnamefont{C.}~\bibnamefont{Cr{\'e}peau}},
  \bibinfo{author}{\bibfnamefont{R.}~\bibnamefont{Jozsa}},
  \bibinfo{author}{\bibfnamefont{A.}~\bibnamefont{Peres}}, \bibnamefont{and}
  \bibinfo{author}{\bibfnamefont{W.~K.} \bibnamefont{Wootters}},
  \bibinfo{journal}{Phys. Rev. Lett.} \textbf{\bibinfo{volume}{70}},
  \bibinfo{pages}{1895} (\bibinfo{year}{1993}).

\bibitem[{\citenamefont{Werner}(2001)}]{werner01}
\bibinfo{author}{\bibfnamefont{R.~F.} \bibnamefont{Werner}},
  \bibinfo{journal}{J. Phys. A: Math. Gen.} \textbf{\bibinfo{volume}{34}},
  \bibinfo{pages}{7081} (\bibinfo{year}{2001}).

\bibitem[{\citenamefont{Vollbrecht and Werner}(2000)}]{vollbrecht00}
\bibinfo{author}{\bibfnamefont{K.~G.~H.} \bibnamefont{Vollbrecht}}
  \bibnamefont{and} \bibinfo{author}{\bibfnamefont{R.~F.}
  \bibnamefont{Werner}}, \bibinfo{journal}{J. Math. Phys.}
  \textbf{\bibinfo{volume}{41}}, \bibinfo{pages}{6772} (\bibinfo{year}{2000}).

\bibitem[{\citenamefont{Horodecki and Horodecki}(1999)}]{horodecki99}
\bibinfo{author}{\bibfnamefont{M.}~\bibnamefont{Horodecki}} \bibnamefont{and}
  \bibinfo{author}{\bibfnamefont{P.}~\bibnamefont{Horodecki}},
  \bibinfo{journal}{Phys. Rev. A} \textbf{\bibinfo{volume}{59}},
  \bibinfo{pages}{4206} (\bibinfo{year}{1999}).

\bibitem[{\citenamefont{Witte and Trucks}(1999)}]{witte99}
\bibinfo{author}{\bibfnamefont{C.}~\bibnamefont{Witte}} \bibnamefont{and}
  \bibinfo{author}{\bibfnamefont{M.}~\bibnamefont{Trucks}},
  \bibinfo{journal}{Phys. Lett. A} \textbf{\bibinfo{volume}{257}},
  \bibinfo{pages}{14} (\bibinfo{year}{1999}).

\bibitem[{\citenamefont{Ozawa}(2000)}]{ozawa00}
\bibinfo{author}{\bibfnamefont{M.}~\bibnamefont{Ozawa}},
  \bibinfo{journal}{Phys. Lett. A} \textbf{\bibinfo{volume}{268}},
  \bibinfo{pages}{158} (\bibinfo{year}{2000}).

\bibitem[{\citenamefont{Bertlmann et~al.}(2005)\citenamefont{Bertlmann,
  Durstberger, Hiesmayr, and Krammer}}]{bertlmann05}
\bibinfo{author}{\bibfnamefont{R.~A.} \bibnamefont{Bertlmann}},
  \bibinfo{author}{\bibfnamefont{K.}~\bibnamefont{Durstberger}},
  \bibinfo{author}{\bibfnamefont{B.~C.} \bibnamefont{Hiesmayr}},
  \bibnamefont{and} \bibinfo{author}{\bibfnamefont{P.}~\bibnamefont{Krammer}},
  \bibinfo{journal}{Phys. Rev. A} \textbf{\bibinfo{volume}{72}},
  \bibinfo{pages}{052331} (\bibinfo{year}{2005}).

\bibitem[{\citenamefont{Vollbrecht and Werner}(2001)}]{vollbrecht01}
\bibinfo{author}{\bibfnamefont{K.~G.~H.} \bibnamefont{Vollbrecht}}
  \bibnamefont{and} \bibinfo{author}{\bibfnamefont{R.~F.}
  \bibnamefont{Werner}}, \bibinfo{journal}{Phys. Rev. A}
  \textbf{\bibinfo{volume}{64}}, \bibinfo{pages}{062307}
  (\bibinfo{year}{2001}).

\bibitem[{\citenamefont{Horodecki and Horodecki}(1996)}]{horodecki96b}
\bibinfo{author}{\bibfnamefont{M.}~\bibnamefont{Horodecki}} \bibnamefont{and}
  \bibinfo{author}{\bibfnamefont{R.}~\bibnamefont{Horodecki}},
  \bibinfo{journal}{Phys. Rev. A} \textbf{\bibinfo{volume}{54}},
  \bibinfo{pages}{1838} (\bibinfo{year}{1996}).

\bibitem[{\citenamefont{Verstraete et~al.}(2002)\citenamefont{Verstraete,
  Audenaert, and Moor}}]{verstraete02}
\bibinfo{author}{\bibfnamefont{F.}~\bibnamefont{Verstraete}},
  \bibinfo{author}{\bibfnamefont{K.}~\bibnamefont{Audenaert}},
  \bibnamefont{and} \bibinfo{author}{\bibfnamefont{B.~D.} \bibnamefont{Moor}},
  \bibinfo{journal}{J. Mod. Opt.} \textbf{\bibinfo{volume}{49}},
  \bibinfo{pages}{1277} (\bibinfo{year}{2002}).

\bibitem[{\citenamefont{Cao and Wang}(2007)}]{cao07}
\bibinfo{author}{\bibfnamefont{Y.}~\bibnamefont{Cao}} \bibnamefont{and}
  \bibinfo{author}{\bibfnamefont{A.~M.} \bibnamefont{Wang}},
  \bibinfo{journal}{J. Phys. A: Math. Theor.} \textbf{\bibinfo{volume}{40}},
  \bibinfo{pages}{3507} (\bibinfo{year}{2007}).

\bibitem[{\citenamefont{Bandyopadhyay et~al.}(2008)\citenamefont{Bandyopadhyay,
  Ghosh, and Roychowdhury}}]{bandyopadhyay08}
\bibinfo{author}{\bibfnamefont{S.}~\bibnamefont{Bandyopadhyay}},
  \bibinfo{author}{\bibfnamefont{S.}~\bibnamefont{Ghosh}}, \bibnamefont{and}
  \bibinfo{author}{\bibfnamefont{V.}~\bibnamefont{Roychowdhury}},
  \bibinfo{journal}{Phys. Rev. A} \textbf{\bibinfo{volume}{77}},
  \bibinfo{pages}{032318} (\bibinfo{year}{2008}).

\bibitem[{\citenamefont{Bertlmann and
  Krammer}(2008{\natexlab{b}})}]{bertlmann08}
\bibinfo{author}{\bibfnamefont{R.~A.} \bibnamefont{Bertlmann}}
  \bibnamefont{and} \bibinfo{author}{\bibfnamefont{P.}~\bibnamefont{Krammer}},
  \bibinfo{journal}{Phys. Rev. A} \textbf{\bibinfo{volume}{77}},
  \bibinfo{pages}{024303} (\bibinfo{year}{2008}{\natexlab{b}}).

\bibitem[{\citenamefont{Bertlmann and
  Krammer}(2008{\natexlab{c}})}]{bertlmann08b}
\bibinfo{author}{\bibfnamefont{R.~A.} \bibnamefont{Bertlmann}}
  \bibnamefont{and} \bibinfo{author}{\bibfnamefont{P.}~\bibnamefont{Krammer}},
  \bibinfo{journal}{Phys. Rev. A} \textbf{\bibinfo{volume}{78}},
  \bibinfo{pages}{014303} (\bibinfo{year}{2008}{\natexlab{c}}).

\bibitem[{\citenamefont{Rudolph}(2000)}]{rudolph00}
\bibinfo{author}{\bibfnamefont{O.}~\bibnamefont{Rudolph}}, \bibinfo{journal}{J.
  Phys. A: Math. Gen.} \textbf{\bibinfo{volume}{33}}, \bibinfo{pages}{3951}
  (\bibinfo{year}{2000}).

\bibitem[{\citenamefont{Rudolph}()}]{rudolph02}
\bibinfo{author}{\bibfnamefont{O.}~\bibnamefont{Rudolph}},
  \bibinfo{note}{e-print arXiv:quant-ph/0202121}.

\bibitem[{\citenamefont{Rudolph}(2003)}]{rudolph03}
\bibinfo{author}{\bibfnamefont{O.}~\bibnamefont{Rudolph}},
  \bibinfo{journal}{Phys. Rev. A} \textbf{\bibinfo{volume}{67}},
  \bibinfo{pages}{032312} (\bibinfo{year}{2003}).

\bibitem[{\citenamefont{Chen and Wu}(2003)}]{chen03}
\bibinfo{author}{\bibfnamefont{K.}~\bibnamefont{Chen}} \bibnamefont{and}
  \bibinfo{author}{\bibfnamefont{L.-A.} \bibnamefont{Wu}},
  \bibinfo{journal}{Quantum Inf. Comput.} \textbf{\bibinfo{volume}{3}},
  \bibinfo{pages}{193} (\bibinfo{year}{2003}).

\bibitem[{\citenamefont{Krammer}(2009)}]{krammer08}
\bibinfo{author}{\bibfnamefont{P.}~\bibnamefont{Krammer}}, \bibinfo{journal}{J.
  Phys. A: Math. Theor.} \textbf{\bibinfo{volume}{42}}, \bibinfo{pages}{065305}
  (\bibinfo{year}{2009}).

\end{thebibliography}

\end{document}